\documentclass[journal]{IEEEtran}
\IEEEoverridecommandlockouts
\usepackage{cite}
\usepackage{amsmath,amssymb,amsfonts}
\usepackage{graphicx}
\usepackage{textcomp}
\usepackage{xcolor}
\usepackage{makecell}
\usepackage{multicol}
\usepackage{siunitx}
\usepackage{booktabs}
\usepackage{xfrac}
\usepackage{algorithm}
\usepackage{algpseudocode}
\usepackage[nolist]{acronym}
\usepackage{tikz}
\usepackage{pgfplots}
\usepackage{rotating}
\pgfplotsset{width=7cm,compat=1.3}
\usetikzlibrary{shapes,arrows,positioning,calc, matrix,spy,patterns}
\usetikzlibrary{decorations.pathreplacing,calligraphy}
\usepgfplotslibrary{groupplots}
\usetikzlibrary{colorbrewer}
\usetikzlibrary{fit}
\usepgfplotslibrary{colorbrewer}
\usepackage{pgfplotstable}
\usepackage{dsfont}%
\newcommand{\mathbbm}[1]{\mathds{#1}}
\usepackage{bm}
\usepackage{dblfloatfix}
\usepackage{mathtools}
\usepackage{upgreek}
\usepackage{algorithm}
\usepackage{algpseudocode}
\usepackage{makecell}
\usepackage{enumitem,kantlipsum}

\definecolor{cb-1}{HTML}{4477AA}
\definecolor{cb-2}{HTML}{EE6677}
\definecolor{cb-3}{HTML}{228833}
\definecolor{cb-4}{HTML}{CCBB44}
\definecolor{cb-5}{HTML}{66CCEE}
\definecolor{cb-6}{HTML}{AA3377}
\definecolor{cb-7}{HTML}{BBBBBB}

\pgfplotscreateplotcyclelist{cb list}{
	cb-1,cb-2,cb-3,cb-4,cb-5,cb-6,cb-7
}

\newcommand{\RevA}[1]{\textcolor{black}{#1}}
\newcommand{\RevB}[1]{\textcolor{black}{#1}}
\definecolor{DarkGreen}{rgb}{0.0, 0.6, 0.0}
\newcommand{\RevC}[1]{\textcolor{black}{#1}}

\usepackage{soul}
\setulcolor{DarkGreen}

\definecolor{kit-green100}{rgb}{0,.59,.51}
\definecolor{kit-green70}{rgb}{.3,.71,.65}
\definecolor{kit-green50}{rgb}{.50,.79,.75}
\definecolor{kit-green30}{rgb}{.69,.87,.85}
\definecolor{kit-green15}{rgb}{.85,.93,.93}
\definecolor{KITgreen}{rgb}{0,.59,.51}

\definecolor{KITpalegreen}{RGB}{130,190,60}
\colorlet{kit-maigreen100}{KITpalegreen}
\colorlet{kit-maigreen70}{KITpalegreen!70}
\colorlet{kit-maigreen50}{KITpalegreen!50}
\colorlet{kit-maigreen30}{KITpalegreen!30}
\colorlet{kit-maigreen15}{KITpalegreen!15}

\definecolor{KITblue}{rgb}{.27,.39,.66}
\definecolor{kit-blue100}{rgb}{.27,.39,.67}
\definecolor{kit-blue70}{rgb}{.49,.57,.76}
\definecolor{kit-blue50}{rgb}{.64,.69,.83}
\definecolor{kit-blue30}{rgb}{.78,.82,.9}
\definecolor{kit-blue15}{rgb}{.89,.91,.95}

\definecolor{KITyellow}{rgb}{.98,.89,0}
\definecolor{kit-yellow100}{cmyk}{0,.05,1,0}
\definecolor{kit-yellow70}{cmyk}{0,.035,.7,0}
\definecolor{kit-yellow50}{cmyk}{0,.025,.5,0}
\definecolor{kit-yellow30}{cmyk}{0,.015,.3,0}
\definecolor{kit-yellow15}{cmyk}{0,.0075,.15,0}

\definecolor{KITorange}{rgb}{.87,.60,.10}
\definecolor{kit-orange100}{cmyk}{0,.45,1,0}
\definecolor{kit-orange70}{cmyk}{0,.315,.7,0}
\definecolor{kit-orange50}{cmyk}{0,.225,.5,0}
\definecolor{kit-orange30}{cmyk}{0,.135,.3,0}
\definecolor{kit-orange15}{cmyk}{0,.0675,.15,0}

\definecolor{KITred}{rgb}{.63,.13,.13}
\definecolor{kit-red100}{cmyk}{.25,1,1,0}
\definecolor{kit-red70}{cmyk}{.175,.7,.7,0}
\definecolor{kit-red50}{cmyk}{.125,.5,.5,0}
\definecolor{kit-red30}{cmyk}{.075,.3,.3,0}
\definecolor{kit-red15}{cmyk}{.0375,.15,.15,0}

\definecolor{KITpurple}{RGB}{160,0,120}
\colorlet{kit-purple100}{KITpurple}
\colorlet{kit-purple70}{KITpurple!70}
\colorlet{kit-purple50}{KITpurple!50}
\colorlet{kit-purple30}{KITpurple!30}
\colorlet{kit-purple15}{KITpurple!15}

\definecolor{KITcyanblue}{RGB}{80,170,230}
\colorlet{kit-cyanblue100}{KITcyanblue}
\colorlet{kit-cyanblue70}{KITcyanblue!70}
\colorlet{kit-cyanblue50}{KITcyanblue!50}
\colorlet{kit-cyanblue30}{KITcyanblue!30}
\colorlet{kit-cyanblue15}{KITcyanblue!15}

\definecolor{KITbraun}{RGB}{167,130,46}

\definecolor{cb-1}{HTML}{4477AA}
\definecolor{cb-2}{HTML}{EE6677}
\definecolor{cb-3}{HTML}{228833}
\definecolor{cb-4}{HTML}{CCBB44}
\definecolor{cb-5}{HTML}{66CCEE}
\definecolor{cb-6}{HTML}{AA3377}
\definecolor{cb-7}{HTML}{BBBBBB}

\pgfplotscreateplotcyclelist{cb list}{
	cb-1,cb-2,cb-3,cb-4,cb-5,cb-6,cb-7
}

\newcommand{\RV}[1]{\mathsf{#1}}

\newcommand*{\vect}[1]{\boldsymbol{#1}}

\newcommand*{\e}{\mathrm{e}}
\let\j\relax
\newcommand{\j}{\mathrm{j}}

\newcommand{\expecv}[2]{\mathbb{E}_{#1} \! \left\{#2\right\}}

\usepackage{xcolor}
\usepackage{amsmath}
\usepackage{amssymb}
\usetikzlibrary{calc}
\usetikzlibrary{positioning}
\usetikzlibrary{patterns}
\usetikzlibrary{decorations.pathreplacing,calligraphy}

\tikzset{
	lsblock/.style={rectangle, thick, draw, minimum width=1.2cm, minimum height=0.6cm, rounded corners=1.6mm, font=\small, align=center}
}

\tikzset{
	aeblock/.style={rectangle, thick, draw, minimum width=2.4cm, minimum height=0.7cm, rounded corners=1.6mm, align=center}
}
\tikzset{
	smlblock/.style={rectangle, thick, draw, minimum width=1.8cm, minimum height=0.75cm, rounded corners=1.2mm, align=center}
}
\tikzset{
	smsblock/.style={rectangle, thick, draw, minimum width=0.6cm, minimum height=0.4cm, rounded corners=1.0mm,inner sep=0.05cm, align=center}
}

\tikzstyle{surround} = [rectangle, rounded corners, draw=KITred, inner sep=0.25cm, dashed, thick]
\tikzstyle{surround_AE} = [rectangle, rounded corners, draw=KITred, inner sep=0.13cm, dashed, thick]

\pgfplotsset{
  colormap={spectral}{
    rgb255=(158,1,66)
    rgb255=(213,62,79)
    rgb255=(244,109,67)
    rgb255=(253,174,97)
    rgb255=(254,224,139)
    rgb255=(230,245,152)
    rgb255=(171,221,164)
    rgb255=(102,194,165)
    rgb255=(50,136,189)
    rgb255=(94,79,162)
  }
}

\pgfplotsset{
  colormap={coolwarm}{
    rgb255=(59,76,192)
    rgb255=(68,90,204)
    rgb255=(77,104,215)
    rgb255=(87,117,225)
    rgb255=(98,130,234)
    rgb255=(108,142,241)
    rgb255=(119,154,247)
    rgb255=(130,165,251)
    rgb255=(141,176,254)
    rgb255=(152,185,255)
    rgb255=(163,194,255)
    rgb255=(174,201,253)
    rgb255=(184,208,249)
    rgb255=(194,213,244)
    rgb255=(204,217,238)
    rgb255=(213,219,230)
    rgb255=(221,221,221)
    rgb255=(229,216,209)
    rgb255=(236,211,197)
    rgb255=(241,204,185)
    rgb255=(245,196,173)
    rgb255=(247,187,160)
    rgb255=(247,177,148)
    rgb255=(247,166,135)
    rgb255=(244,154,123)
    rgb255=(241,141,111)
    rgb255=(236,127,99)
    rgb255=(229,112,88)
    rgb255=(222,96,77)
    rgb255=(213,80,66)
    rgb255=(203,62,56)
    rgb255=(192,40,47)
    rgb255=(180,4,38)
  }
}

\pgfplotsset{
    sensing colormap/.style={
      colormap name=spectral,     %
      point meta=z,
      point meta min=0,         %
      point meta max= 80,         %
    }
  }

  \pgfplotsset{
    communication colormap/.style={
      colormap name=coolwarm,     %
      point meta min=2,         %
      point meta max= 4.6,         %
    }
  }

\def\BibTeX{{\rm B\kern-.05em{\sc i\kern-.025em b}\kern-.08em
    T\kern-.1667em\lower.7ex\hbox{E}\kern-.125emX}}

\begin{document}
\begin{acronym}[]
    \acro{5G}{fifth generation}
    \acro{6G}{sixth generation}
    \acro{AE}{autoencoder}
    \acro{AIR}{achievable information rate}
    \acro{AWG}{arbitrary waveform generator}
    \acro{AWGN}{additive white Gaussian noise}
    \acro{CP}{cyclic prefix}
    \acro{CPAS}{conventional probabilistic amplitude shaping}
    \acro{CRB}{Cramér-Rao bound}
    \acro{DFT}{discrete Fourier transform}
    \acro{DR}{dynamic range}
    \acro{EISL}{expected integrated sidelobe level}
    \acro{GMI}{generalized mutual information}
    \acro{GMM}{Gaussian mixture model}
    \acro{GPAS}{generalized probabilistic amplitude shaping}
    \acro{IDFT}{inverse discrete Fourier transform}
    \acro{ISAC}{integrated sensing and communications}
    \acro{LLR}{log-likelihood ratio}
    \acro{LTI}{linear-time invariant}
    \acro{MBA}{modified Blahut-Arimoto}
    \acro{MF}{matched filter}
    \acro{MI}{mutual information}
    \acro{ML}{maximum-likelihood}
    \acro{MMSE}{minimum mean square error}
    \acro{MSE}{mean squared error}
    \acro{NN}{neural network}
    \acro{OFDM}{orthogonal frequency division multiplexing}
    \acro{PAPR}{peak-to-average-power ratio}
    \acro{PAS}{probabilistic amplitude shaping}
    \acro{PDF}{probability density function}
    \acro{PSK}{phase shift keying}
    \acro{RCS}{radar cross section}
    \acro{RDM}{range-Doppler matrix}
    \acro{QAM}{quadrature amplitude modulation}
    \acro{QPSK}{quadrature phase shift keying}
    \acro{RF}{reciprocal filter}
    \acro{RV}{random variable}
    \acro{SC}[S\&C]{sensing \& communications}
    \acro{SCE}{sensing channel estimator}
    \acro{ISINR}{image signal-to-interference-and-noise ratio}
    \acro{SNR}{signal-to-noise ratio}
    \acro{TOI}{target of interest}
    \acro{WF}{Wiener filter}
\end{acronym}

\bstctlcite{IEEEexample:BSTcontrol}

\title{Constellation Shaping for OFDM-ISAC Systems:\\A Systematic and Experimental Comparison}

\author{
Benedikt Geiger,~\IEEEmembership{Graduate Student Member, IEEE}, Daniel Gil Gaviria,~\IEEEmembership{Graduate Student Member, IEEE},\\ Christian Karle,~\IEEEmembership{Graduate Student Member, IEEE}, Delcho Donev,\\Charlotte Muth,~\IEEEmembership{Graduate Student Member, IEEE}, Thomas Zwick,~\IEEEmembership{Fellow, IEEE},\\ Jürgen Becker,~\IEEEmembership{Senior Member, IEEE}, Benjamin Nuss,~\IEEEmembership{Senior Member, IEEE}, Laurent Schmalen,~\IEEEmembership{Fellow, IEEE}
\thanks{This work has received funding from the German Federal Ministry of Research, Technology and Space (BMFTR) within the projects Open6GHub, Open6GHub+, and KOMSENS-6G (grant agreements 16KISK010, 16KIS2405, and 16KISK123).}%
\thanks{Benedikt Geiger, Daniel Gil Gaviria, Charlotte Muth, and Laurent Schmalen are with the Communications Engineering Lab (CEL), Karlsruhe Institute of Technology (KIT), Hertzstr. 16, 76187 Karlsruhe, Germany (e-mail: first.lastname@kit.edu). Christian Karle and Jürgen Becker are with the Institute for Information Processing Technology (ITIV), KIT, 76131 Karlsruhe, Germany. Delcho Donev is with Rohde \& Schwarz GmbH \& Co. KG, 81671 Munich, Germany. Thomas Zwick is and Benjamin Nuss was with the Institute of Radio Frequency Engineering and Electronics (IHE), KIT, 76131 Karlsruhe, Germany. Benjamin Nuss is now with the Professorship of Microwave Sensors and Sensor Systems, Technical University of Munich (TUM), 80333 Munich, Germany.}
}

\maketitle

\begin{abstract}
Emerging \ac{ISAC} systems reuse the communications signal for radar-like sensing to provide environmental awareness. However, this integrated design imposes conflicting constraints on the modulation format, as communications systems use higher-order modulation formats while sensing systems use unit-modulus signals, making constellation shaping a key performance driver in \ac{ISAC} systems. This paper develops a unified analytical and experimental framework for constellation shaping in monostatic orthogonal frequency division multiplexing \ac{ISAC} systems and systematically analyzes its interaction with the three canonical \acp{SCE}: the matched filter, the reciprocal filter, and the \ac{WF}. We begin by deriving constellation-dependent \ac{ISINR} expressions for the \acp{SCE} that account for practical effects such as windowing and multi target scenarios. Based on this, we derive the corresponding entropy-maximizing channel input distributions. 
We generalize an autoencoder framework to optimize geometric, probabilistic, including conventional and generalized low-complexity probabilistic amplitude shaping (PAS), and joint constellation shaping across all three \acp{SCE}. Extensive simulations and hardware-in-the-loop experiments validate the derived \ac{ISINR} and show that sensing performance is governed by the \ac{SCE} choice, while communications performance is dominated by the shaping method. Notably, the \ac{WF} combined with joint shaping achieves the best \ac{ISAC} trade-off. Low-complexity PAS variants closely approach this performance, offering a flexible trade-off and outperforming legacy modulation formats.
\end{abstract}


\acresetall
\vspace{-0.7cm}
\section{Introduction}
\vspace{-0.15cm}
Over the past decade, communications and radar systems have shown a growing convergence in their requirements, including wider bandwidths, higher carrier frequencies, larger antenna arrays, and the adoption of \ac{OFDM} as a common waveform~\cite{liu_integrated_2022}. Therefore, next-generation communications systems are envisioned to integrate sensing capabilities by reusing the \ac{OFDM} communications signal in a radar-like manner~\cite{wild_6g_2023,liu_integrated_2022}, improving energy, hardware, and spectral efficiency compared to operating two separate systems.

However, the constellation choice introduces a fundamental trade-off in these \ac{ISAC} systems. While communications systems employ high-order \ac{QAM} to increase spectral efficiency, radar systems favor unit-modulus constellations, as higher-order \ac{QAM} degrades sensing performance~\cite{keskin_fundamental_2025}.

\RevA{This degradation depends on the \ac{SCE} employed at the sensing receiver to remove the information-bearing transmit symbols. Common \acp{SCE} include the \ac{MF}, which maximizes the output \ac{SNR} at the cost of signaling-induced interference, the \ac{RF}, which avoids this interference but introduces noise enhancement, and the \ac{WF}, which balances both effects by minimizing the channel estimation \ac{MSE}, while attenuating the target peak~\cite{keskin_fundamental_2025}. Accordingly, different intermediate metrics have been used to characterize the individual degradation mechanisms. For example, the \ac{EISL} quantifies signaling-induced sidelobes but does not account for noise enhancement or the attenuation of the target peak. In contrast, the output \ac{SNR} captures the target power relative to noise but neglects signaling-induced interference. While the \mbox{\ac{MSE}} captures target attenuation, noise enhancement, and signaling-induced interference, it describes them as estimation error and does not compare the power of the target to the resulting image noise-and-interference \mbox{floor~\cite{du_probabilistic_2025}}. This limits its ability to compare the considered \mbox{\acp{SCE}~\cite{du_probabilistic_2025}}.}

\setulcolor{blue}\RevC{While~\cite{han_constellation_2025} derives the \ac{MSE} of delay estimation for the \ac{MF} and \ac{RF}, extending this analysis to the \ac{WF} is not straightforward, since the resulting expressions cannot be simplified in the same manner. Furthermore, \mbox{although~\cite{du_probabilistic_2025}} starts from a multi target signal model, its subsequent image domain \mbox{\acf{DR}} analysis focuses on an isolated target. However, for the \mbox{\ac{MF}} and \mbox{\ac{WF}}, additional targets increase the signaling-induced interference, such that the analysis \mbox{in~\cite{du_probabilistic_2025}} does not characterize the resulting multi target behavior. Moreover, practical receiver-side windowing is not included in the existing analytical expressions. Consequently, experimental verification of these \mbox{\ac{SCE}}-dependent effects has largely remained qualitative, e.g., through range or Doppler cuts\mbox{~\cite{du_probabilistic_2025,xu_experimental_2024,han_constellation_2025}}, whereas a systematic quantitative comparison between theory and measurements across different \mbox{\acp{SCE}} is still missing. Moreover, post-\mbox{\ac{SCE}} noise statistics have received little attention.}

\setulcolor{red}\RevC{To enable a direct comparison between the \mbox{\ac{MF}}, \mbox{\ac{RF}}, and \mbox{\ac{WF}} and to provide a consistent basis for optimization and quantitative experimental validation, it is necessary to jointly incorporate these practical effects into a single, unified metric.} \RevA{At the same time, the resulting \ac{ISINR} provides a more generally applicable intermediate characterization than task-specific end-to-end metrics, such as the detection probability, whose evaluation requires additional assumptions, e.g., on the detector and false-alarm probability.}

\begin{figure*}[!t]
    \centering
    \input{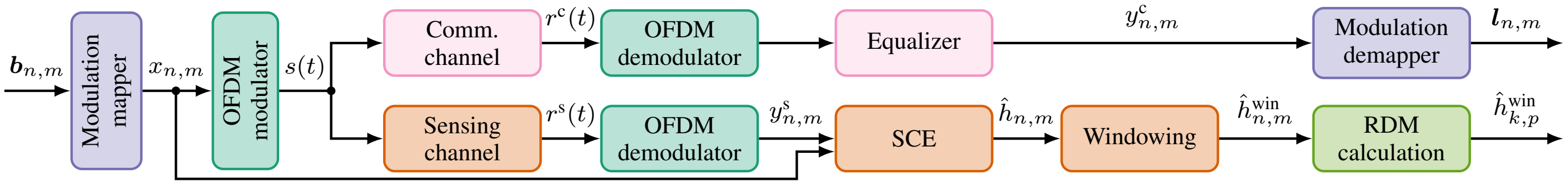}
    \vspace{-0.5cm}
    \caption{Block diagram of the considered monostatic OFDM-ISAC system. The transmit signal is for both \ac{SC} tasks.}
    \label{fig:system_model:block_diagram}
    \vspace{-0.4cm}
\end{figure*}

Since the modulation format fundamentally determines both \ac{SC} performance~\cite{xiong_torch_2024}, constellation shaping has emerged as a practical tool to govern the \ac{SC} trade-off inherent in \ac{ISAC} systems and to outperform legacy modulation formats such as \ac{QAM}. Constellation shaping can be broadly categorized into (i) geometric shaping, which optimizes the constellation symbol locations assuming equal probabilities of occurrence; (ii) probabilistic shaping, which optimizes the probabilities of occurrence of the symbols under fixed constellation symbol locations (e.g., for \ac{QAM}); and (iii) joint shaping, which jointly optimizes both the symbol locations and their probabilities of occurrence.

While \ac{AE}-based constellation optimization is well established in communications~\cite{stark_joint_2019,rode_end--end_2023}, its extension to \ac{ISAC} is recent. Prior work has studied probabilistic shaping for the \ac{MF} and compared geometric, probabilistic, and joint shaping, showing that joint shaping achieves the best overall \ac{SC} performance~\cite{du_probabilistic_2023,du_reshaping_2023,Geiger25JCS}. Further, low-implementation complexity \ac{CPAS} and \ac{GPAS} enable near-joint without receiver-side error propagation~\cite{geiger_constellation_2025}.

\setulcolor{blue}\RevC{So far, constellations optimized for \mbox{\ac{ISAC}~\cite{du_probabilistic_2025,geiger_constellation_2025}} are designed for specific sensing constraints. However, practical \mbox{\ac{ISAC}} systems must dynamically adapt their modulation to varying \mbox{\ac{SC}} requirements. While a codebook of offline-optimized constellations may be stored for a discretized set of \mbox{\ac{SC}} requirements, a fine sensing constraint grid increases the memory requirements, whereas coarse quantization may either sacrifice communications performance or violate the sensing requirement and online training is computationally impractical. This motivates continuously adaptive \mbox{constellations~\cite{cammerer_trainable_2020}} for \mbox{\ac{ISAC}} that can generate a constellation for the currently required sensing constraint without online retraining or the storage of a densely sampled constellation codebook. For the \mbox{\ac{WF}}, such an adaptation is particularly relevant since the constellation additionally depends on the sensing scenario.}

\setulcolor{DarkGreen}
Since the \ac{SCE} determines how signal randomness translates into sensing performance degradation, the constellations have to be designed for each \mbox{\ac{SCE}} \mbox{independently~\cite{han_constellation_2025, du_probabilistic_2025}}. \RevB{So far, only specific combinations have been explored, such as \mbox{geometric~\cite{han_constellation_2025}} or \mbox{probabilistic~\cite{du_probabilistic_2025, yang_constellation_2024}} shaping for the \mbox{\ac{RF}} or probabilistic shaping for the \mbox{\ac{WF}~\cite{du_probabilistic_2025}}. In particular, \mbox{\cite{du_probabilistic_2025}} fixes the constellation points and restricts the design to probabilistic shaping, while employing an \mbox{\ac{MSE}}-based sensing constraint to obtain a convex optimization problem. Hence, neither is a comparable image domain metric optimized directly, nor is the \mbox{\ac{SC}} trade-off achievable by jointly optimizing geometric and probabilistic shaping explored. Therefore, the literature lacks a unified comparison of geometric, probabilistic, and joint shaping across different \mbox{\acp{SCE}}.}

Moreover, while low-complexity implementations based on \ac{CPAS} and \ac{GPAS} achieve near-joint-shaping performance for the \ac{MF}, it is currently unknown whether this result generalizes to the \ac{RF} and the \ac{WF}. Finally, prior work on the \ac{RF} and \ac{WF} has predominantly focused on the evaluation of sensing performance~\cite{keskin_fundamental_2025, du_probabilistic_2025}, while the communications performance has largely been overlooked, leaving the \ac{SC} performance trade-off insufficiently characterized.

While individual aspects of constellation shaping for \ac{ISAC} have been studied previously, this paper provides an analytical and experimental framework for constellation shaping in \ac{OFDM}-\ac{ISAC} systems, enabling a systematic comparison of geometric, probabilistic, and joint shaping across all common \acp{SCE} under a single \ac{ISINR} framework. The main contributions of this paper are: %
\begin{enumerate}[wide, labelindent=0pt]
    \vspace{-0.1cm}
    \item[\textbf{1)}] \setulcolor{DarkGreen}\RevA{\mbox{\textbf{Common analytical sensing metric:}} We derive closed form \mbox{\ac{ISINR}} expressions for the \mbox{\ac{MF}}, \mbox{\ac{RF}}, and \mbox{\ac{WF}}, accounting for signaling-induced sidelobes, noise enhancement, target peak attenuation, practical windowing, and multi target interference. This provides the analytical basis for the subsequent optimization and a common intermediate metric for quantitative comparison and validation. We further relate the \mbox{\ac{ISINR}} to end-to-end sensing performance, e.g., the detection probability.}
    \item[\textbf{2)}] \textbf{Estimator-specific analytical characterization:} Based on the derived \ac{ISINR}, we analyze how different \acp{SCE} handle transmit signal randomness, providing analytical insights for the optimization, and derive the post-\ac{SCE} noise statistics.
    \item[\textbf{3)}]\RevC{\textbf{Theoretical channel input \ac{PDF} design under sensing constraints:} To analyze the impact of \ac{SCE}-specific sensing constraints, we derive the continuous, entropy-maximizing input \acp{PDF} for the \ac{MF} and \ac{RF} in closed form. This provides a fundamental theoretical reference that complements the numerical constellation optimization.}
    \item[\textbf{4)}]\RevB{\textbf{Unified autoencoder-based constellation optimization:} We generalize the \mbox{\ac{AE}} framework \mbox{of~\cite{geiger_constellation_2025}} along three main dimensions: from the \mbox{\ac{MF}} only to a systematic comparison across the \mbox{\acp{SCE}}, from a single \mbox{\ac{SNR}} to multi \mbox{\ac{SNR}} generalization, and from individually optimized fixed constellations to a continuously adaptive constellation family. Compared \mbox{with~\cite{du_probabilistic_2025}}, we optimize geometric, probabilistic, and joint constellation shaping, including low-complexity \mbox{\ac{CPAS}} and \mbox{\ac{GPAS}}, under a comparable \mbox{\ac{ISINR}} constraint. Compared with the convex formulation \mbox{in~\cite{du_probabilistic_2025}}, the enlarged design space together with the \mbox{\ac{ISINR}} constraint results in a non-convex optimization problem but also an improved \mbox{\ac{SC}} trade-off.}
    \item[\textbf{5)}]\textbf{Comprehensive experimental validation:} We conduct extensive hardware-in-the-loop experiments and simulations to validate the derived \ac{ISINR} and constellation shaping gains, demonstrating how the \ac{SCE} choice and constellation shaping jointly determine the achievable \ac{SC} trade-off in practical \ac{OFDM}-\ac{ISAC} systems.
\end{enumerate}

\vspace{-0.1cm}
\textit{Notations:} Bold lowercase letters (e.g., $\vect{x}$) denote vectors, bold uppercase letters matrices (e.g., $\vect{J}$), and scalars appear in normal font (e.g., $N$). Sans-serif symbols (e.g.,~$\RV{x}$) denote random variables. The index indicates the domain: $t$ continuous time, $k$ and $p$ discrete delay and Doppler, and $m$ and $n$ the OFDM symbol and subcarrier index. The functions $\mathbbm{h}(\cdot)$ and $\mathbb{H}(\cdot)$ denote differential and discrete entropy.

\section{System Model}
\label{sec:ch2:system_model}

\subsection{\acs{ISAC} Transmitter and Constellation Shaping}
\RevC{We consider the monostatic \ac{OFDM}-\ac{ISAC} system shown in Fig.~\ref{fig:system_model:block_diagram}, where the sensing receiver is co-located with the \ac{ISAC} transmitter and a single transmit signal is used for \ac{SC}.} The transmitter generates an \ac{OFDM} frame consisting of $M$ \ac{OFDM} symbols indexed by \mbox{$m \in \{0,\ldots,M-1\}$}, each comprising $N$ subcarriers with subcarrier spacing $\Delta f$, indexed by $n \in \{0,\ldots,N-1\}$. 
For each resource element, the constellation mapper maps $Q$ bits \mbox{$\bm{b}_{n,m} \in \{0,1\}^{Q}$} onto a complex modulation symbol \mbox{$x_{n,m} \in \mathcal{X}$}, where \mbox{$\mathcal{X} \subset \mathbb{C}$} denotes the modulation alphabet with cardinality \mbox{$|\mathcal{X}| \eqqcolon \tilde{Q} = 2^Q$}. The same constellation mapping is used across the entire frame, such that the transmit symbols $x_{n,m}$ are modeled as independent and identically distributed \acp{RV}~\mbox{$\RV{x}_{n,m}$} following the distribution $P_{\RV{x}}(x)$ for all $n$ and $m$. Throughout this work, the transmit symbols are assumed to be zero mean and normalized to unit average power, i.e.,
\vspace{-0.1cm}
\begin{equation}
    \expecv{\RV{x}}{\RV{x}_{n,m}} = 0, \qquad \expecv{\RV{x}}{|\RV{x}_{n,m}|^2} = 1, \qquad \forall n,m.
    \label{eq:system_model:zero_mean_unit_power}
    \vspace{-0.05cm}
\end{equation}
Geometric, probabilistic, or joint constellation shaping is obtained through the choice of $\mathcal{X}$ and $P_{\RV{x}}(x)$, where \ac{CPAS} and \ac{GPAS} realize low-complexity probabilistic shaping by constraining $P_{\RV{x}}(x)$ to factorize into real and imaginary components or into a radial distribution with uniform phase, respectively~\cite{geiger_constellation_2025}. The \ac{OFDM} modulator transforms each \ac{OFDM} symbol into the time domain via an orthonormal \ac{IDFT} and appends a \ac{CP}, resulting in a total \ac{OFDM} symbol duration of $T_{\text{sym}}$. Digital-to-analog conversion yields the equivalent continuous-time baseband signal $s(t)$.

\subsection{Communications Channel, Receiver, and Performance Metric}
\vspace{-0.15cm}
For communications, we consider a deterministic but arbitrary \ac{LTI} channel $h^{\text{c}}(t)$ impaired by \ac{AWGN} $\RV{w}^{\text{c}}(t)$, represented by \mbox{$\RV{r}^{\text{c}}(t) = h^{\text{c}}(t) * \RV{s}(t) + \RV{w}^{\text{c}}(t)$}.%

The communications receiver samples the baseband signal $\RV{r}^{\text{c}}(t)$, removes the \ac{CP}, and converts each received \ac{OFDM} symbol into the frequency domain using an orthonormal \ac{DFT}. After linear \ac{MMSE} channel equalization, the \ac{OFDM} system can be modeled as a set of $N$ parallel \ac{AWGN} channels, one for each subcarrier, with channel transition \ac{PDF} $f_{\RV{y}^{\text{c}}_n|\RV{x}_n}(y^{\text{c}}_{n,m}|x_{n,m})$. The received communications symbol on the $n$th subcarrier and $m$th \ac{OFDM} symbol after equalization can therefore be modeled as
\vspace{-0.1cm}
\begin{equation}
    \RV{y}^{\text{c}}_{n,m} = \RV{x}_{n,m} + \RV{w}^{\text{c}}_{n,m},
    \label{eq:system_model:comm_channel}
    \vspace{-0.05cm}
\end{equation}
where $\RV{w}^{\text{c}}_{n,m} \sim \mathcal{CN}(0, \sigma_{\text{c},n}^2)$ is \ac{AWGN} with variance $\sigma_{\text{c},n}^2$.

A bitwise demapper computes the \ac{LLR} for each received symbol and bit position (see, e.g.~\cite{Ivanov_BICM})
\begin{equation}
    l_{n,m,q}(y^{\text{c}}_{n,m}) \! = \!  \log \frac{\sum_{x \in \mathcal{X}^{(0)}_q}{ \! f_{\RV{y}^{\text{c}}_n|\RV{x}_n}(y^{\text{c}}_{n,m}|x)}P_{\RV{x}}(x)}{\sum_{x \in \mathcal{X}^{(1)}_q}{ \! f_{\RV{y}^{\text{c}}_n|\RV{x}_n}(y^{\text{c}}_{n,m}|x)}P_{\RV{x}}(x)},
    \label{eq:system_model:LLR}
\end{equation}
where $\mathcal{X}^{(b)}_q$ represents the set of all constellation symbols labeled with bit \mbox{$b \in \{0,1\}$} at bit position $q$.

To evaluate the communications performance achieved in practical
communications systems with bit metric decoding, we use the \ac{GMI}~\cite{bocherer_bandwidth_2015,fehenberger_probabilistic_2016}
\begin{equation}
    \mathbb{I}_{\text{GMI}}(\bm{\RV{b}}_{n,m}; \RV{y}^{\text{c}}_{n,m}) \!=\!\! \left[ \mathbb{H}(\bm{\RV{b}}_{n,m}) \!-\!\! \sum\limits_{q=1}^{Q} \mathbb{H}(\RV{b}_{n,m,q}|\RV{y}^{\text{c}}_{n,m}) \!\right]^{\!+}\!\!\!\!, %
    \label{eq:system_model:GMI}
\end{equation}
where $[\; \cdot\;  ]^+$ is $\mathrm{max}(\cdot,\,0)$. The \ac{GMI}~(\ref{eq:system_model:GMI}) gives a lower bound on the \ac{AIR}.

\subsection{Sensing Channel and Receiver, and Performance Metric}
Assuming an arbitrary but fixed sensing channel with $J$~point targets, each having distance $R_j$, radial velocity $v_j$, and \ac{RCS} $\sigma_{\text{RCS,}j}$, the received sensing signal is\cite{braun_ofdm_2014}
\begin{equation}
    \RV{r}^{\text{s}}(t) = \sum_{j=1}^{J} a_j \RV{s}(t - \tau_j) \e^{\j ( 2\pi f_{\text{D,}j} t + \varphi_j)}+ \RV{w}^{\text{s}}(t),
    \label{eq:system_model:sensing_channel_time_domain}
\end{equation}
where \mbox{$\RV{w}^{\text{s}}(t) \! \sim \! \mathcal{CN} \left( 0, \sigma_{\text{s}}^2 \right)$} denotes \ac{AWGN} with variance $\sigma_{\text{s}}^2$ and the phases~$\varphi_j$ are drawn uniformly from \mbox{$[0,2\pi)$} and are treated as fixed parameters thereafter. The amplitude $a_j$, delay $\tau_j$, and Doppler frequency $f_{\text{D},j}$ of target $j$ are \cite{braun_ofdm_2014}
\begin{equation}
    a_j\! = \! \sqrt{ \frac{\sigma_{\text{RCS},j} c_0^2 P_{\text{Tx}} G_{\text{Tx}} G_{\text{Rx}}}{\left( 4 \pi \right)^3 R_j^4 f_{\text{c}}^2}}, \,
    \tau_j \! = \! 2 \frac{R_j}{c_0}, \,
    f_{\text{D,}j} \! = \! 2 \frac{v_j f_{\text{c}}}{c_0},
\end{equation}
where $P_{\text{Tx}}$, $G_{\text{Tx}}$, $G_{\text{Rx}}$, $f_{\text{c}}$, and $c_0$ denote transmit power, gain of the transmit and receive antenna, carrier frequency, and the speed of light, respectively. \setulcolor{DarkGreen}\RevA{We use the term \textit{target} for any modeled echo component for notational convenience. Such components may physically originate from object scattering centers, discrete clutter, or multipath-induced reflections~\mbox{\cite{kumari_ieee_2018}}. Spatially extended objects can be represented by multiple point scatterers in this framework.}

The sensing receiver samples the baseband signal $\RV{r}^{\text{s}}(t)$, removes the \ac{CP}, and transforms each of the $M$ received \ac{OFDM} symbols into the frequency domain using the orthonormal \ac{DFT}. 
The resulting received symbols in the frequency domain are \mbox{$\RV{y}^{\text{s}}_{n,m} = h_{n,m} \RV{x}_{n,m} + \RV{w}^{\text{s}}_{n,m}$}, where the doubly selective sensing channel is~\cite{braun_ofdm_2014}
\begin{equation}
    h_{n,m} = \sum_{j=1}^{J} a_j \e^{\j \varphi_j} \e ^{- \j 2 \pi \frac{n}{N} \tilde{\tau}_j } \e^{\j 2 \pi  \frac{m}{M} \tilde{f}_{\text{D,}j}},
    \label{eq:system_model:sensing_channel_frequency_domain}
\end{equation}
with normalized delay \mbox{$\tilde{\tau}_j = \Delta f \tau_j N$} and Doppler \mbox{$\tilde{f}_{\text{D,}j} = T_{\text{sym}} f_{\text{D,}j}M$}, and $\RV{w}^{\text{s}}_{n,m}$ is the \ac{DFT} of the sampled \ac{AWGN} $\RV{w}^{\text{s}}(t)$. %

The \ac{SCE} $g_{n,m}$, which can be either an \ac{MF}, \ac{RF}, or \ac{WF}, is defined as~\cite{keskin_fundamental_2025}
\begin{equation}
    g_{n,m} = \begin{cases}
        x^\star_{n,m}, & \text{MF}, \\
        \frac{1}{x_{n,m}}, & \text{RF}, \\
        \frac{x^{\star}_{n,m}}{|x_{n,m}|^2 + \tilde{\sigma}_{\text{s}}^2}, & \text{WF},
    \end{cases}
    \vspace{-0.15cm}
\end{equation}
with $\tilde{\sigma}_{\text{s}}^2 = \sigma^2_{\text{s}}/\sum_{j=1}^{J}a^2_j$ denoting the sensing noise variance normalized to the total target power. The \ac{SCE} tries to eliminate the transmit symbols $\RV{x}_{n,m}$ from the received sensing symbols $\RV{y}^{\text{s}}_{n,m}$, yielding the frequency-domain sensing channel estimate
\vspace{-0.1cm}
\begin{equation}
    \hat{\RV{h}}_{n,m} = \RV{y}^{\text{s}}_{n,m} \cdot \RV{g}_{n,m}.
    \label{eq:system_model:frequency_domain_channel_estimate}
    \vspace{-0.1cm}
\end{equation}

To suppress sidelobes in the sensing image, a separable time-frequency window $u_{n,m}$, e.g., a Hamming window, is applied to the channel estimate along both the subcarrier and \ac{OFDM} symbol dimensions
\vspace{-0.1cm}
\begin{equation}
    \hat{\RV{h}}^{\text{win}}_{n,m} = \hat{\RV{h}}_{n,m} \cdot u_{n,m}.
    \label{eq:system_model:windowing}
    \vspace{-0.1cm}
\end{equation}
The window is normalized to unit power, i.e., \mbox{$\frac{1}{MN}\sum_{m=0}^{M-1}\sum_{n=0}^{N-1}u^2_{n,m} = 1$}. 

An oversampled delay-Doppler representation (sensing image or \ac{RDM}) $\hat{\RV{h}}^{\text{win}}_{k,p}$ is obtained by evaluating the orthonormal \ac{IDFT} across the subcarriers and the orthonormal \ac{DFT} across the \ac{OFDM} symbols on an oversampled delay-Doppler grid
\vspace{-0.2cm}
\begin{equation}
    \hat{\RV{h}}^{\text{win}}_{k,p}= \frac{1}{\sqrt{MN}} \sum_{m=0}^{M-1} \sum_{n=0}^{N-1} \hat{\RV{h}}^{\text{win}}_{n,m} \mathrm{e}^{\mathrm{j} 2 \pi \frac{n}{\eta_\tau N} k} \mathrm{e}^{-\mathrm{j} 2 \pi \frac{m}{\eta_\text{D} M} p},
    \label{eq:system_model:RDM_calculation}
    \vspace{-0.05cm}
\end{equation}
with \mbox{$k = \{0,\dots,\eta_\tau N-1\}$} and \mbox{$p = \{0,\dots,\eta_\text{D} M-1\}$}, where $\eta_\tau$ and $\eta_\text{D}$ denote the oversampling factors in delay and Doppler direction, respectively.

\RevA{We build on the mean and variance decomposition for data-dependent sensing channel estimates~\cite{liu_uncovering_2025} and the image-domain noise-and-interference observations in~\cite{nuss_limitations_2018}. Motivated by these observations, we define the \ac{ISINR} for the \ac{TOI} as}
\begin{equation}
    \mathrm{ISINR}_{\text{TOI}}
    =
    \frac{\left| \mathbb{E}_{\RV{x},\RV{w}^{\text{s}}} \{ \hat{\RV{h}}^{\text{win}} _{k_{\text{TOI}}, p_{\text{TOI}} } \}  \right|^2}
    {\frac{1}{\eta_{\text{D}}M\eta_\tau N} \sum_{p=0}^{\eta_{\text{D}}M-1}\sum_{k=0}^{\eta_\tau N-1} \mathrm{Var}_{\RV{x},\RV{w}^{\text{s}}} \! \left\{ \hat{\RV{h}}^{\text{win}} _{k, p}  \right\}},
    \label{eq:system_model:ISINR}
\end{equation}
which assumes that the \ac{TOI} with index~$j_{\text{TOI}}$ is well separated from all remaining targets, i.e.,
\vspace{-0.1cm}
\begin{equation}
    | R_{j_{\text{TOI}}} - R_j | \!\gg\! \Delta R,
    \quad \! \!
    \left|v_{j_{\text{TOI}}}  - v_{j} \right| \! \gg \! \Delta v,
    \quad \!\! 
    \forall j \neq j_{\text{TOI}},
    \label{eq:theory:target_separation_condition}
    \vspace{-0.05cm}
\end{equation}
with range resolution \mbox{$\Delta R = \frac{c_0}{2 N \Delta f}$} and velocity resolution \mbox{$\Delta v = \frac{c_0}{2 f_{\text{c}} M T_{\text{sym}}}$}.
\RevA{The \ac{TOI} denotes the target whose sensing performance is evaluated and is predefined during system evaluation, i.e., simulations, and measurements, \mbox{analogous} to e.g., the use of ground-truth target parameters for evaluating the \ac{MSE} of range estimation.}

\setulcolor{red}\RevB{The \mbox{\ac{ISINR}} is defined as the ratio between the deterministic \mbox{\ac{TOI}} contribution and the surrounding noise-and-interference floor. In contrast, the \mbox{\ac{DR}} or peak-to-sidelobe metric used in concurrent \mbox{work~\cite{keskin_fundamental_2025,du_probabilistic_2025}} compares the total power in the \mbox{\ac{TOI}} bin, including noise and interference, with the asymptotic image floor. The \mbox{\ac{DR}} therefore directly characterizes the observable peak-to-floor ratio in the sensing image. In contrast, the \mbox{\ac{ISINR}} isolates the deterministic target contribution and can be related directly to end-to-end sensing performance, as described in \mbox{Sec.~\ref{sec:theory:e2e}}. For an approximately homogeneous noise-and-interference floor and statistically independent targets, the two metrics are related by \mbox{$\mathrm{DR}_{\text{TOI}} \approx 1+\mathrm{ISINR}_{\text{TOI}}$} and therefore become approximately equivalent in the high \mbox{\ac{ISINR}} regime.}\footnote{\RevB{However, the difference between the \mbox{\ac{DR}} and \mbox{\ac{ISINR}} is crucial at low \mbox{\ac{ISINR}}, where detection performance is critical and substituting the \mbox{\ac{DR}} for the \mbox{\ac{ISINR}} \mbox{in~\eqref{eq:pd_isinr}} would overestimate the corresponding detection probability.}}

\RevA{Note that the present \ac{ISINR} formulation considers only a two-dimensional \ac{RDM}. Conceptually, it can be extended to an array-based range-Doppler-angle representation by defining the \ac{TOI} at a specific range-Doppler-angle cell and including angular sidelobe leakage in the interference term. For a fixed array and beamforming configuration, the same principle therefore remains applicable. A comprehensive system optimization would, however, require the joint consideration of constellation shaping, array geometry, spatial power allocation, and transmit/receive beamforming, which is beyond the scope of this work and left for future research.}

\section{Derivation of Constellation- and SCE-Dependent ISINR}
\label{sec:theory:SCE}
In this section, we derive and analyze how the random transmit symbols and the \ac{SCE} jointly determine the sensing performance, establishing the analytical basis for the subsequent constellation optimization.

\begin{table*}[]
\vspace{-0.2cm}
\caption{Scaling factor $\alpha$, sensing-channel estimate variance, and resulting \ac{ISINR} for the considered \acp{SCE}}
\vspace{-0.25cm}
    \centering
    \normalsize
    \begin{tabular}{c c c c c}
        \toprule
        \ac{SCE} & \ac{MF} & \ac{RF} & \ac{WF} \\
        \midrule
        Scaling factor $\alpha$ & $1$ & $1$ & $\mathbb{E}_{\RV{x}} \left\{\frac{|\RV{x}|^2}{|\RV{x}|^2+\tilde{\sigma}_{\text{s}}^2}\right\}$ \\
        Variance $\mathrm{Var}^{\text{win}}_{\RV{x},\RV{w}^{\text{s}}} \{ \hat{\RV{h}}_{k,p} \}$ & $\sigma_{\text{s}}^2 + (\expecv{\RV{x}}{|\RV{x}|^4}-1) \sum_{j=1}^{J} a^2_j$ & $\sigma_{\text{s}}^2\expecv{\RV{x}}{|\RV{x}|^{-2}}$ & $\sigma_{\text{s}}^2 \expecv{\RV{x}}{ \! \frac{|\RV{x}|^2}{ \left( |\RV{x}|^2 \! + \! \tilde{\sigma}_{\text{s}}^2 \right)^2} \!}  + \mathrm{Var}_{\RV{x}} \! \left\{ \frac{|\RV{x}|^2}{ |\RV{x}|^2 \!+\! \tilde{\sigma}_{\text{s}}^2 }\right\} \sum_{j=1}^{J} a^2_j $ \\
        $\mathrm{ISINR}_{\text{TOI}}$ & $\frac{a^2_{j_\text{TOI}} G_{\text{P}}}{ \sigma_{\text{s}}^2 + (\expecv{\RV{x}}{|\RV{x}|^4}-1) \sum_{j=1}^{J} a^2_j }$ & $\frac{a^2_{j_\text{TOI}} G_{\text{P}}}{ \sigma_{\text{s}}^2\expecv{\RV{x}}{|\RV{x}|^{-2}}}$ & $\frac{a^2_{j_\text{TOI}} \mathbb{E}^2_{\RV{x}} \left\{\frac{|\RV{x}|^2}{|\RV{x}|^2+\tilde{\sigma}_{\text{s}}^2}\right\} G_{\text{P}}}{\sigma_{\text{s}}^2 \expecv{\RV{x}}{ \! \frac{|\RV{x}|^2}{ \left( |\RV{x}|^2 \! + \! \tilde{\sigma}_{\text{s}}^2 \right)^2} \!}  + \mathrm{Var}_{\RV{x}} \left\{ \frac{|\RV{x}|^2}{ |\RV{x}|^2 \!+\! \tilde{\sigma}_{\text{s}}^2 }\right\} \sum_{j=1}^{J} a^2_j}$ \\
        \bottomrule
    \end{tabular}
    \label{tab:ch3:Image_SINR}
    \vspace{-0.6cm}
\end{table*}

\subsection{From Frequency-domain Channel Estimate to ISINR}
The target peak in the \ac{RDM} originates from the deterministic sensing channel response~\eqref{eq:system_model:sensing_channel_frequency_domain}, while the noise-and-interference floor in the \ac{RDM} is caused by the noise of the sensing channel and the randomness of the transmit symbols. Therefore, we decompose the frequency-domain sensing channel estimate for all three considered \acp{SCE} as
\vspace{-0.15cm}
\begin{align}
    \hat{\RV{h}}_{n,m} &= \RV{y}^{\text{s}}_{n,m}\cdot \RV{g}_{n,m} = (\RV{x}_{n,m} h_{n,m} + \RV{w}^{\text{s}}_{n,m})\cdot \RV{g}_{n,m} \notag \\&=\RV{x}_{n,m}\RV{g}_{n,m} h_{n,m} \!+\! \RV{w}^{\text{s}}_{n,m} \RV{g}_{n,m} \! + \! \alpha h_{n,m} \! - \! \alpha h_{n,m} \label{eq:theory:channel_estiate_seperability} \\ & = \!\!\!\!\!\!\!\!\!\!\!\!\!\!\!\!\!\underbrace{\alpha h_{n,m}}_{\text{Deterministic target response}} \!\!\!\!\!\!\!\!\!\!\! + \ \ \ \ \underbrace{h_{n,m}(\RV{x}_{n,m} \RV{g}_{n,m}-\alpha) + \RV{w}^{\text{s}}_{n,m} \RV{g}_{n,m}}_{\text{Noise and random signaling } \tilde{\RV{w}}^{\text{s}}_{n,m}} \notag
\end{align} %
where \mbox{$\alpha = \expecv{\RV{x}}{\RV{x}_{n,m} \cdot \RV{g}_{n,m}}$} is an \ac{SCE}-dependent real-valued scaling factor and $\tilde{\RV{w}}^{\mathrm{s}}_{n,m}$ is zero-mean and uncorrelated. By construction, the variance of $\tilde{\RV{w}}^{\text{s}}_{n,m}$ equals that of the channel estimate, i.e., \mbox{$\mathrm{Var}_{\RV{x},\RV{w}^{\text{s}}}\{\tilde{\RV{w}}^{\text{s}}_{n,m}\} = \mathrm{Var}_{\RV{x},\RV{w}^{\text{s}}}\{\hat{\RV{h}}_{n,m}\}$}.

This decomposition enables an independent analysis of the deterministic target contribution and the noise-and-interference floor. Consequently, the \ac{ISINR} is governed by how these two components propagate through the two-dimensional unitary \ac{DFT} and \ac{IDFT} operations used to form the \ac{RDM}, as well as the receiver-side windowing operation~\eqref{eq:system_model:windowing}.

For a well-separated \ac{TOI}, the deterministic target component coherently adds up at the target bin $(k_{\text{TOI}},p_{\text{TOI}})$ in the delay-Doppler domain, yielding the processing gain
\vspace{-0.2cm}
\begin{equation}
    G_{\text{P}} \coloneqq \underbrace{NM}_{\text{Coherent integration gain}} \underbrace{\left( \frac{1}{NM} \sum_{m=0}^{M-1} \sum_{n=0}^{N-1} \!u_{n,m} \right)^2}_{\text{Windowing loss}},
    \label{eq:theory:SCE:processing_gain}
    \vspace{-0.25cm}
\end{equation}
where the factor $NM$ corresponds to the coherent integration gain, while deviations from a rectangular window introduce a windowing loss. Finally, the deterministic power of the peak of the \ac{TOI} is
\vspace{-0.5cm}
\begin{equation}
    \left| \expecv{\RV{x},\RV{w}^{\text{s}}}{\hat{\RV{h}}^{\text{win}}_{k_{\text{TOI}},p_{\text{TOI}}}} \right|^2 \approx  \alpha^2 a^2_{j_\text{TOI}} G_{\text{P}}. 
    \label{eq:theory:target_power}
    \vspace{-0.2cm}
\end{equation}
\textit{Proof:} See Appendix~\ref{ch:appendix:target_power}.

For the noise-and-interference power, the use of orthonormal \ac{DFT} and \ac{IDFT} operations and a unit average power window $u_{n,m}$ implies that the variance of the delay-Doppler channel estimate approximately equals the averaged variance in the time-frequency domain
\vspace{-0.2cm}
\begin{equation}
    \mathrm{Var}_{\RV{x},\RV{w}^{\text{s}}} \! \left\{ \hat{\RV{h}}^{\text{win}}_{k,p} \right\} \approx \frac{1}{NM} \sum_{m=0}^{M-1} \sum_{n=0}^{N-1} \mathrm{Var}_{\RV{x},\RV{w}^{\text{s}}}\! \left\{ \hat{\RV{h}}_{n,m} \right\}.
    \label{eq:theory:noise-and-interference-power}
    \vspace{-0.2cm}
\end{equation}
\textit{Proof:} See Appendix~\ref{ch:appendix:noise-and-interference-power}.

Consequently, the \ac{ISINR} for the \ac{TOI} is
\vspace{-0.25cm}
\begin{equation}
    \mathrm{ISINR}_{\text{TOI}}
    =
    \frac{
        \alpha^2 a^2_{j_{\text{TOI}}} \, G_{\mathrm{P}}
    }{
        \frac{1}{MN}\sum_{m=0}^{M-1}\sum_{n=0}^{N-1}\mathrm{Var}_{\RV{x},\RV{w}^{\text{s}}}\{\hat{\RV{h}}_{n,m}\}
    }.
    \label{eq:theory:ISINR}
    \vspace{-0.15cm}
\end{equation}

\noindent\textbf{Remark:}
For the considered \acp{SCE}, the \ac{ISINR} is fully determined by the first- and second-order moments of the frequency domain sensing channel estimate, namely its mean (through $\alpha$) and variance $\mathrm{Var}\{\hat{\RV{h}}_{n,m}\}$. Windowing reduces the deterministic target power, but not the noise-and-interference floor.

\vspace{-0.4cm}
\subsection{Mean and Variance of \acp{SCE}}
\vspace{-0.15cm}
We now discuss the mean and variance of the sensing channel estimates for the considered \acp{SCE}, which are summarized in Table~\ref{tab:ch3:Image_SINR}. The scaling factor $\alpha$ (mean) directly follows from its definition. The variance of the channel estimate $\mathrm{Var}_{\RV{x},\RV{w}^{\text{s}}}\{ \hat{\RV{h}}^{\text{win}}_{k,p} \}$ in the delay-Doppler domain is obtained by first calculating the variance $\mathrm{Var}_{\RV{x},\RV{w}^{\text{s}}}\{ \hat{\RV{h}}_{n,m}\}$ in the frequency domain and inserting it into~(\ref{eq:theory:noise-and-interference-power}), where we exploit that \mbox{$\frac{1}{MN}\sum_m \sum_n |h_{n,m}|^2 = \sum_j a_j^2$}. The derivation builds on~\cite{keskin_fundamental_2025, du_probabilistic_2025} but extends their results to multi target scenarios considered in this work. Therefore, the intermediate steps are omitted for brevity.

\subsubsection{Reciprocal filter (RF)} removes the transmit symbols through symbol-wise division
\label{sec:theory:SCE:mean_var:RF}
\vspace{-0.5cm}
\begin{equation}
    \hat{\RV{h}}^{\text{RF}}_{n,m} = \frac{\RV{y}^{\text{s}}_{n,m}}{\RV{x}_{n,m}} = h_{n,m} + \overbrace{\underbrace{\frac{\RV{w}^{\text{s}}_{n,m}}{\RV{x}_{n,m}},}_{\tilde{\RV{w}}^{\text{s}}_{n,m}}}^{\text{Noise enhancement}}
    \label{eq:system_model:definition:SD}
    \vspace{-0.3cm}
\end{equation}
which yields an unbiased sensing channel estimate, i.e., \mbox{$\alpha^{\text{RF}} = 1$}, but scales the noise with the inverse of the transmit symbol $\frac{1}{\RV{x}_{n,m}}$ resulting in noise enhancement. Conditioned on a transmit symbol \mbox{$x_{\tilde{q}} = \RV{x}_{n,m}$}, the noise is Gaussian distributed with variance $\sigma^2_{\text{s}}/|x_{\tilde{q}}|^2$. Marginalizing over the transmit symbols leads to a \ac{GMM} as \ac{PDF} 
\vspace{-0.2cm}
\begin{equation}
    f_{\tilde{\RV{w}}^{\text{s}}}(\tilde{w}^{\text{s}}) = \sum_{{\tilde{q}}=1}^{\tilde{Q}} P_{\RV{x}}(x_{\tilde{q}}) \frac{|x_{\tilde{q}}|^2}{\pi \sigma^2_{\text{s}}} \exp \left( - |\tilde{w}^{\text{s}}|^2 \frac{|x_{\tilde{q}}|^2}{\sigma^2_{\text{s}}}\right).
\end{equation}
As a result, small amplitude constellation points cause noise enhancement, increasing the variance of $\hat{\RV{h}}^{\text{RF}}_{n,m}$ and $\hat{\RV{h}}^{\text{RF,win}}_{k,p}$, which scales with the inverse second-order moment of the transmit symbols (Tab.~\ref{tab:ch3:Image_SINR}),
\vspace{-0.2cm}
\begin{equation}
    \mathrm{Var}_{\RV{x},\RV{w}^{\text{s}}}\left\{ \hat{\RV{h}}^{\text{RF,win}}_{k,p} \right\} = \sigma^2_{\text{s}} \cdot \expecv{\RV{x}}{|\RV{x}|^{-2}}.
    \label{eq:theory:SCE:RF:noise_enhancement}
\end{equation} %

\subsubsection{Matched filter (MF)} multiplies the received symbol with the complex-conjugate transmit symbol
\vspace{-0.3cm}
\begin{equation}
    \begin{split}
        \hat{\RV{h}}^{\text{MF}}_{n,m} & = \RV{y}^{\text{s}}_{n,m}\RV{x}_{n,m}^{\star} = h_{n,m} |\RV{x}_{n,m}|^{2} + \RV{w}^{\text{s}}_{n,m} \RV{x}_{n,m}^{\star} \\
        & = h_{n,m} + \underbrace{\overbrace{h_{n,m}(|\RV{x}_{n,m}|^{2} - 1)}^{\text{Interference}} + \RV{w}^{\text{s}}_{n,m} \RV{x}_{n,m}^{\star}}_{\tilde{\RV{w}}^{\text{s}}_{n,m}}.
    \end{split}
    \label{eq:system_model:definition:MF}
    \vspace{-0.2cm}
\end{equation}
yielding an unbiased channel estimate, i.e., \mbox{$\alpha^{\text{MF}} = 1$}. 
The \ac{MF} maximizes the output \ac{SNR} for known $\RV{x}_{n,m}$. The deterministic target response is scaled by the instantaneous symbol power $|\RV{x}_{n,m}|^2$, resulting in signaling-induced interference that manifests as sidelobes. The variance of the \ac{MF} channel estimate
\vspace{-0.3cm}
\begin{equation}
    \mathrm{Var}_{\RV{x},\RV{w}^{\text{s}}} \{\hat{\RV{h}}^{\text{MF,win}}_{k,p}\}  = \sigma_{\text{s}}^2 + (\expecv{\RV{x}}{|\RV{x}|^4}-1) \sum_{j=1}^{J} |a_j|^2
    \vspace{-0.2cm}
\end{equation}
scales with both the aggregate target power and the fourth moment of the constellation. \setulcolor{red}\RevC{Consequently, clutter and multipath components also increase the signaling-induced interference besides forming additional peaks in the \mbox{\ac{RDM}}.} For zero mean, unit average power constellations~(\ref{eq:system_model:zero_mean_unit_power}), the fourth moment corresponds to the kurtosis \mbox{$\kappa = \expecv{\RV{x}}{|\RV{x}|^4}$}.

Moreover, under the unit average power constraint, the kurtosis is directly related to the variance of the symbol power, \mbox{$\expecv{\RV{x}}{|\RV{x}|^4} = \mathrm{Var}_{\RV{x}}\! \{|\RV{x}|^2\} + 1$}, and thus quantifies the fluctuation of the symbol power around the unit circle~\cite{geiger_constellation_2025}.

In the high-SNR sensing regime, i.e., when the sensing noise variance is much smaller than the aggregate target power (\mbox{$\tilde{\sigma}_{\text{s}}^2 \ll 1$}), target-induced interference causes the \ac{ISINR} to saturate at
\vspace{-0.3cm}
\begin{equation}
    \mathrm{ISINR}_{\text{TOI}} = \frac{a^2_{j_\text{TOI}} G_{\text{P}}}{(\expecv{\RV{x}}{|\RV{x}|^4}\!-\!1) \sum_{j=1}^{J} a^2_j } \overset{\text{TOI}}{\underset{\text{only}}{=}} \frac{G_{\text{P}}}{\expecv{\RV{x}}{|\RV{x}|^4}\!-\!1}. %
    \label{eq:theory:SCE:MF:Image_SINR_saturation}
    \vspace{-0.25cm}
\end{equation}

In contrast, in the low-SNR sensing regime (\mbox{$\tilde{\sigma}_{\text{s}}^2 \gg 1$}), the interference is negligible, and, following the same reasoning as in the \ac{RF} case, the \ac{MF} scales the noise by the complex-conjugate transmit symbol $x^\star_{\tilde{q}}$ resulting in a variance $\sigma^2_{\text{s}} |x_{\tilde{q}}|^2$. Marginalizing over the transmit symbols yields the sensing noise \ac{PDF} 
\vspace{-0.3cm}
\begin{equation}
f_{\tilde{\RV{w}}^{\text{s}}}(\tilde{w}^{\text{s}}) = \sum_{\tilde{q}=1}^{\tilde{Q}}
P_{\RV{x}}(x_{\tilde{q}})\,
\frac{1}{\pi\sigma_{\text{s}}^{2}|x_{\tilde{q}}|^{2}}
\exp\!\left(-\frac{|\tilde{w}^{\text{s}}|^{2}}{\sigma_{\text{s}}^{2}|x_{\tilde{q}}|^{2}}\right).
\vspace{-0.15cm}
\end{equation}

\begin{figure}
    \centering
    \input{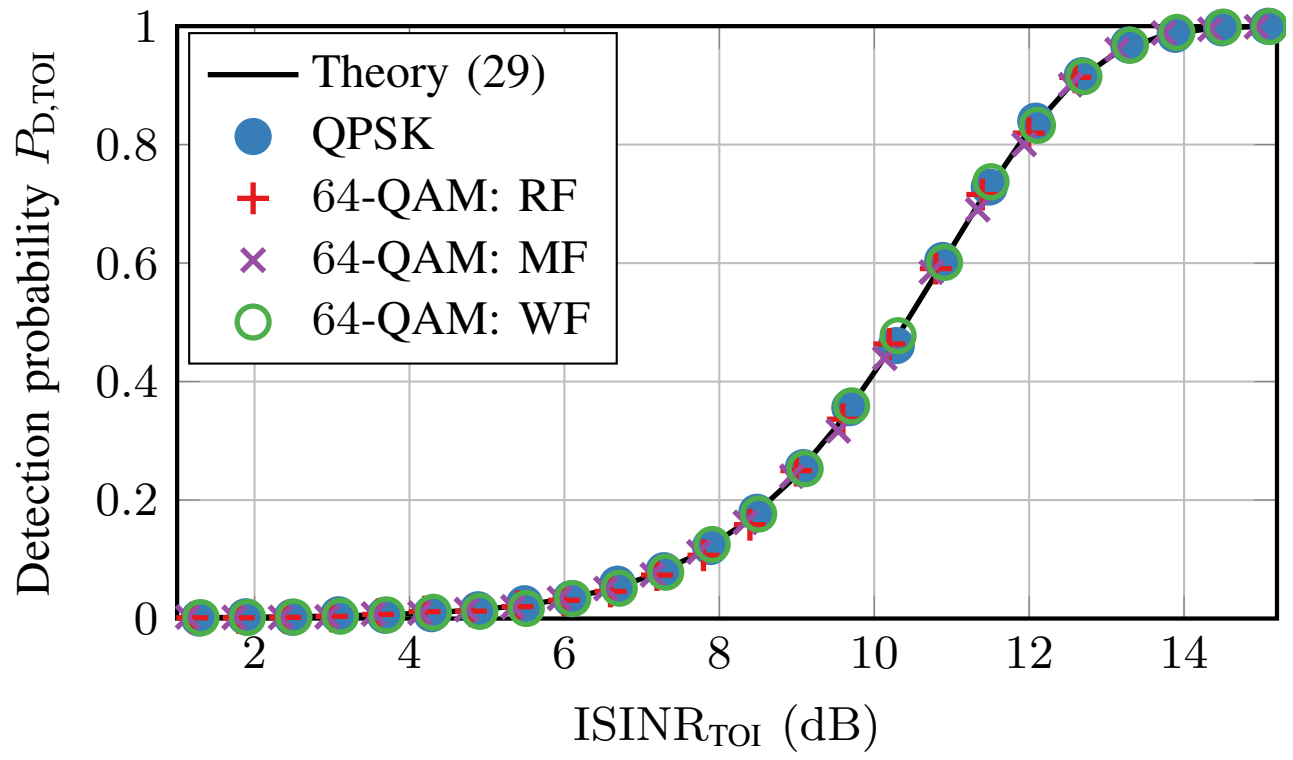}
    \vspace{-0.5cm}
    \caption{Detection probability versus \ac{ISINR}: comparison of the theory with simulations using \ac{QPSK} and \num{64}-\ac{QAM} with different \acp{SCE}. A two target scenario with well separated, on-grid targets is considered, where a weak \ac{TOI} is detected in the presence of a substantially stronger interfering target.}
    \label{fig:theory:ISINR_detection_prob}
    \vspace{-0.5cm}
\end{figure}

\subsubsection{Wiener filter (WF)} generalizes the \ac{MF} and \ac{RF} by performing a linear \ac{MMSE} estimate of the sensing channel, where the normalized sensing noise variance regularizes the inversion of the transmit symbol.
\vspace{-0.2cm}
\begin{equation}
    \begin{split}
       & \hat{\RV{h}}^{\text{WF}}_{n,m} =  \frac{\RV{y}^{\text{s}}_{n,m}\RV{x}_{n,m}^{\star}}{|\RV{x}_{n,m}|^{2} \!\!+\! \tilde{\sigma}_{\text{s}}^{2}} \!=\! \frac{h_{n,m}|\RV{x}_{n,m}|^{2} \!\!+\! \RV{w}^{\text{s}}_{n,m}\RV{x}_{n,m}^{\star}}{|\RV{x}_{n,m}|^{2} \!\!+\! \tilde{\sigma}_{\text{s}}^{2}} \\
       = & \!\!\!\!\! \overbrace{\alpha^{\text{WF}}}^{\text{Attenuation}} \!\!\!\!\!\! h_{n,m} \!\!+\! \underbrace{\overbrace{h_{n,m}\! \!\left(\! \! \frac{|\RV{x}_{n,m}|^2}{|\RV{x}_{n,m}|^{2} \!\!+\! \tilde{\sigma}_{\text{s}}^{2}} \!-\! \alpha^{\text{WF}}\! \! \right)}^{\text{Interference}} \!\!+\!\! \overbrace{\frac{\RV{w}^{\text{s}}_{n,m}\RV{x}_{n,m}^{\star}}{|\RV{x}_{n,m}|^{2} \!\! +\! \tilde{\sigma}_{\text{s}}^{2}}}^{\text{Noise enhancement}}}_{\tilde{\RV{w}}^{\text{s}}_{n,m}}\!,
       \label{eq:system_model:definition:WF}
    \end{split}
\end{equation}
where the sensing channel estimate is attenuated by
\vspace{-0.1cm}
\begin{equation}
    \vspace{-0.1cm}
    \alpha^{\text{WF}} = \mathbb{E}_{\RV{x}} \left\{\frac{|\RV{x}|^2}{|\RV{x}|^2+\tilde{\sigma}_{\text{s}}^2}\right\},
    \vspace{-0.05cm}
\end{equation}
yielding a biased channel estimate. The \ac{WF} smoothly transitions between the \ac{RF} behavior in the high sensing \ac{SNR} regime and the \ac{MF} behavior in the low sensing \ac{SNR} regime, introducing a combination of noise enhancement and interference. Consequently, its variance
\vspace{-0.2cm}
\begin{equation}
    \mathrm{Var}_{\RV{x},\RV{w}^{\text{s}}}\!\!\left\{ \!\hat{\RV{h}}^{\text{WF,win}}_{k,p} \! \right\} \!=\! \sigma_{\text{s}}^2 \expecv{\RV{x}}{\! \! \frac{|\RV{x}|^2}{ \left( |\RV{x}|^2 \! + \! \tilde{\sigma}_{\text{s}}^2 \right)^2} \!\!}  \!+\!\! \mathrm{Var}_{\RV{x}}\! \! \left\{\! \frac{|\RV{x}|^2}{ |\RV{x}|^2 \!+\! \tilde{\sigma}_{\text{s}}^2 } \!\!\right\} \! \sum_{j=1}^{J} \!a^2_j
\end{equation}
contains a noise enhancement term, which dominates in the high sensing \ac{SNR} regime, and an interference term, which dominates in the low sensing \ac{SNR} regime. 

Inserting the scaling factors $\alpha$ and variances into~(\ref{eq:system_model:ISINR}) yields the \ac{ISINR} expressions for all \acp{SCE}, summarized in Tab.~\ref{tab:ch3:Image_SINR}, which are the basis for the following optimization.

\vspace{-0.8cm}
\subsection{From ISINR to End-to-End Sensing Performance}\label{sec:theory:e2e}
\RevA{The \ac{ISINR} serves as a common analytical intermediate metric that can be related to task-specific sensing performance once the corresponding detector or estimator is specified. In the following, we illustrate this relation for target detection. Similar relations for the \ac{MSE} of delay or Doppler estimation can be established following~\cite{richards_fundamentals_2014,braun_ofdm_2014,han_constellation_2025}.}

\RevA{For detection, we consider a square-law detector. Due to the superposition of many data-dependent interference and noise contributions in the \ac{RDM}, the noise and interference can be approximated as circularly symmetric complex Gaussian in the \ac{RDM} according to the central limit theorem~\cite{geiger_constellation_2025}. Under the considered system model, detection therefore reduces to detecting a deterministic \ac{TOI} in additive Gaussian noise. For a given false-alarm probability~$P_{\text{FA}}$, the corresponding detection probability is~\cite{richards_fundamentals_2014}}
\RevA{\begin{equation}
    P_{\text{D,TOI}} = Q_1\left(
        \sqrt{2\cdot\mathrm{ISINR}_{\text{TOI}}},
        \sqrt{-2\ln P_{\text{FA}}}
    \right),
    \label{eq:pd_isinr}
\end{equation}}
\noindent\RevA{where \(Q_1(\cdot,\cdot)\) denotes the first-order Marcum \(Q\)-function. Hence, for a fixed \(P_{\text{FA}}\), the detection probability increases monotonically with the \ac{ISINR}, as confirmed in Fig.~\ref{fig:theory:ISINR_detection_prob}, where we verify~\eqref{eq:pd_isinr} with various numerical simulations.}

\vspace{0.1cm}
\section{Constellation Shaping Under Sensing Constraints}
\label{sec:optimization}
In this section, we formulate the constellation design problem with a unified sensing constraint, derive entropy-maximizing channel input \acp{PDF} under the resulting sensing constraints, and introduce the \ac{AE} framework for constellation optimization.

\begin{table}[!t]
\caption{\ac{SCE}-dependent sensing metric $\Gamma(\mathcal{X}, P_{\RV{x}}(x))$ and corresponding sweep ranges from the sensing-optimal to the communications-optimal point.}
\vspace{-0.2cm}
    \centering
    \normalsize
    \begin{tabular}{c c c}
        \toprule
        SCE & Sensing metric $\Gamma(\mathcal{X}, P_{\RV{x}}(x))$ & Range\\
        \midrule
        \ac{MF} & $\expecv{\RV{x}}{|\RV{x}|^4}$ & $[1,2]$\\
        \ac{RF} & $\expecv{\RV{x}}{|\RV{x}|^{-2}}$ & $[1, \infty)$\\
        \ac{WF} & $\frac{\expecv{\RV{x}}{ \! \frac{|\RV{x}|^2}{ \left( |\RV{x}|^2 \! + \! \tilde{\sigma}_{\text{s}}^2 \right)^2} \!}  + \mathrm{Var}_{\RV{x}}\! \left\{ \frac{|\RV{x}|^2}{ |\RV{x}|^2 \!+\! \tilde{\sigma}_{\text{s}}^2 }\right\} / \tilde{\sigma}^2_{\text{s}}}{\mathbb{E}^2_{\RV{x}} \left\{\frac{|\RV{x}|^2}{|\RV{x}|^2+\tilde{\sigma}_{\text{s}}^2}\right\}}$ & $[1, \infty)$ \\
        \bottomrule
    \end{tabular}
    \label{tab:theory:Sensing_constraint}
    \vspace{-0.5cm}
\end{table}

\vspace{-0.3cm}
\subsection{Sensing Metric and Optimization Problem}
\vspace{-0.1cm}
We aim to find constellations tailored to each \ac{SCE} that maximize the communications performance of an \ac{OFDM}-\ac{ISAC} system, measured by the \ac{GMI}, while ensuring a predefined sensing performance in terms of \ac{ISINR}.

The analysis in the previous section showed that the \ac{ISINR}, given in Tab.~\ref{tab:ch3:Image_SINR}, depends on both the sensing channel and the employed \ac{SCE}, such that directly constraining the \ac{ISINR} introduces an explicit dependence on the sensing channel. To decouple constellation design from sensing channel realizations, we introduce a scalar sensing metric~$\Gamma(\mathcal{X}, P_{\RV{x}}(x))$, which removes as many channel-dependent terms as possible and isolates the constellation-dependent contribution to the \ac{ISINR}. The explicit expression of the sensing metric~$\Gamma(\mathcal{X}, P_{\RV{x}}(x))$ depends on the \ac{SCE} and is summarized in Table~\ref{tab:theory:Sensing_constraint}. The key idea is that $\Gamma(\mathcal{X}, P_{\RV{x}}(x))$ represents the constellation-intrinsic component of the \ac{ISINR}. It is monotonically related to the \ac{ISINR}, with smaller $\Gamma(\mathcal{X}, P_{\RV{x}}(x))$ corresponding to higher \ac{ISINR}. For the \ac{MF} and \ac{RF}, the sensing metric $\Gamma(\mathcal{X}, P_{\RV{x}}(x))$ depends solely on the statistical distribution of the transmit constellation, whereas for the \ac{WF}, the sensing metric $\Gamma(\mathcal{X}, P_{\RV{x}}(x))$ additionally depends on the normalized sensing noise power $\tilde{\sigma}^2_{\text{s}}$.

To jointly optimize \ac{SC} performance, we maximize the \ac{GMI} under the sensing constraint \mbox{$\Gamma(\mathcal{X}, P_{\RV{x}}(x)) \leqslant \tilde{\Gamma}$}, leading to the following optimization problem~\cite{geiger_constellation_2025}
\vspace{-0.2cm}
\begin{align}
    \max_{\mathcal{X}, P_{\RV{x}}(x)} \ & \sum_{m=0}^{M-1} \sum_{n=0}^{N-1} \mathbb I_{\text{GMI} } \left( \RV{x}_{n,m}; \RV{y}^{\text{c}}_{n,m}\right) \label{eq:theory:optimization_problem:GMI_opt} \\
    \text{s.t. } & \textstyle \! \sum_{x}P_{\RV{x}}(x) = 1, \quad P_{\RV{x}}(x) \geqslant 0, \quad \forall x \in \mathcal{X}, \tag{C\num{0}} \label{eq:theory:optimization_problem:C0} \\
    & \mathbb{E}_{\RV{x}} \{ | \RV{x} |^2 \} = 1, \quad \; \; \mathbb{E}_{\RV{x}} \{ \RV{x} \} = 0, \tag{C\num{1}} \label{eq:theory:optimization_problem:C1} \\
    & \Gamma(\mathcal{X}, P_{\RV{x}}(x)) \leqslant \tilde{\Gamma} .\tag{C\num{2}} \label{eq:theory:optimization_problem:C2}
    \vspace{-0.15cm}
\end{align}
Here, the constraints~(\ref{eq:theory:optimization_problem:C0}) enforce that $P_{\RV{x}}(x)$ satisfies the properties of a probability mass function, while the constraints~(\ref{eq:theory:optimization_problem:C1}) ensure that the constellation has unit average power and zero mean. \RevB{Note that the optimization problem~\eqref{eq:theory:optimization_problem:GMI_opt} is generally non-convex~\cite{steiner_comparison_2017}, since optimizing the \ac{GMI} with respect to the constellation points is non-convex and the \ac{ISINR}-based sensing constraints are not guaranteed to define a convex feasible set under joint geometric and probabilistic shaping.}

To characterize the \ac{SC} trade-off, we sweep the threshold~$\tilde{\Gamma}$ between the two extreme points of the range of $\Gamma(\mathcal{X}, P_{\RV{x}}(x))$, listed in Tab.~\ref{tab:theory:Sensing_constraint}. Increasing the threshold~$\tilde{\Gamma}$ relaxes the sensing requirement, enlarges the feasible constellation set, and enables a continuous transition from sensing-optimal to communications-optimal constellation shapes.

Despite these \ac{SCE}-dependent definitions, the optimal constellations at the extreme points of the \ac{SC} trade-off are identical for all \acp{SCE}. Unit-modulus constellations are sensing-optimal and maximize the \ac{ISINR}, for which all \acp{SCE} coincide up to the constant scaling factor $\alpha$. Conversely, a circularly symmetric complex Gaussian \ac{PDF} maximizes the \ac{MI} of an \ac{AWGN} channel under an average-power constraint, independent of the \ac{SCE}.

\vspace{-0.3cm}
\subsection{\RevC{Theoretical Considerations: Entropy Maximization under Sensing Constraints}}
\RevC{The constrained \ac{GMI} maximization problem in the previous subsection does not admit a closed-form solution in general. However, for sufficiently high \ac{SNR}, the bit-wise conditional entropy after demapping becomes negligible, such that maximizing the \ac{GMI} reduces to maximizing the input entropy}
\vspace{-0.3cm}
\begin{equation}
    \mathbb I_{\text{GMI}}(\bm{\RV{b}}; \RV{y}^{\text{c}}) 
    = \mathbb H(\RV{x}) 
    - \underbrace{\sum_{q=1}^{Q} \mathbb H(\RV{b}_{q}|\RV{y}^{\text{c}})}_{\rightarrow\,0}.
    \vspace{-0.2cm}
\end{equation}
\RevC{While \ac{GMI} optimization at finite \ac{SNR} yields discrete constellation points, probabilities, and bit labels for practical transmission systems, entropy maximization provides a continuous theoretical reference and structural insight into the resulting input distribution. We therefore relax the discrete constellation to a continuous input and maximize the differential entropy $\mathbbm{h}(\RV{x})$ subject to \ac{PDF} normalization, unit average power, and the corresponding \ac{SCE}-dependent moment constraint. Introducing Lagrange multipliers and applying the stationarity condition of the resulting variational problem shows that the optimal unique global maximum-entropy density belongs to the exponential family~\cite[Ch.~12.1]{Cover2006}, since the differential entropy is strictly concave. Focusing on the \ac{MF} and \ac{RF} as the two fundamental constraint types, the \ac{RF} constrains the inverse second-, zeroth-, and second-order moments, whereas the \ac{MF} constrains the zeroth-, second-, and fourth-order moments. The corresponding entropy-maximizing input \acp{PDF} are}
\vspace{-0.15cm}
\begin{equation}
    f_{\RV{x}_{\text{max}}}(x) = \begin{cases}
    \e^{\gamma^{\text{RF}}_{-2}/|x|^2 + \gamma^{\text{RF}}_{0} + \gamma^{\text{RF}}_{2} |x|^2}, & \text{RF}, \\
    \e^{\gamma^{\text{MF}}_{0} + \gamma^{\text{MF}}_{2} |x|^2 + \gamma^{\text{MF}}_{4} |x|^4}, & \text{MF}.
    \end{cases}
    \label{eq:Bound:Ansatz_Maximum_entropy}
    \vspace{-0.15cm}
\end{equation}

The real-valued parameters $\gamma$ are determined such that the corresponding moment constraints are satisfied. Specifically, the constraints can be written as 
\vspace{-0.15cm}
\begin{equation}
    C_{2c} = \begin{cases}
        \int_{\mathbb{C}} |x|^{2c} \e^{\gamma^{\text{RF}}_{-2}/|x|^2 + \gamma^{\text{RF}}_{0} + \gamma^{\text{RF}}_{2} |x|^2} \, \mathrm{d}x, & \text{RF}, \\
        \int_{\mathbb{C}} |x|^{2c} \e^{\gamma^{\text{MF}}_{0} + \gamma^{\text{MF}}_{2} |x|^2 + \gamma^{\text{MF}}_{4} |x|^4} \, \mathrm{d}x, & \text{MF},  \end{cases}
    \label{eq:bound:non_linear_eq_system}  
    \vspace{-0.15cm}
\end{equation}
where \mbox{$c \in \{-1,0,1\}$} for the \ac{RF} and \mbox{$c \in \{0,1,2\}$} for the \ac{MF} respectively. Here, \mbox{$C_0 = C_2 = 1$} enforce \ac{PDF} normalization and the unit average power constraint, whereas $C_{-2}$ and $C_4$ capture the \ac{SCE}-dependent sensing constraints.

In both cases, the resulting \acp{PDF} depend on $|x|$ only and are therefore circularly symmetric, implying zero mean. The constraint equations form a system of nonlinear equations involving integrals over the complex plane. While this system can be solved numerically in general, Appendix~\ref{sec:appendix:entropy_maximization} shows that, for the \ac{RF}, it can be reduced to the equivalent set of equations
\vspace{-0.6cm}
\begin{align}
    C_{-2} & = 2 \pi \e^{\gamma_0} K_0\left( 2 \sqrt{\gamma_2 \gamma_{-2}}\right), \label{eq:theory:Cm2} \\
    C_0 & = 2 \pi \e^{\gamma_0} \sqrt{\frac{\gamma_{-2}}{\gamma_2}} K_1\left( 2 \sqrt{\gamma_2 \gamma_{-2}}\right), \label{eq:theory:C0} \\
    C_2 & = 2 \pi \e^{\gamma_0} \frac{\gamma_{-2}}{\gamma_2} K_2\left( 2 \sqrt{\gamma_2 \gamma_{-2}}\right), \label{eq:theory:C2}
\end{align}
which can be solved more efficiently since no integrals over the complex plane are required. Here $K_\nu(\cdot)$ is the modified Bessel function of order $\nu$. An analogous simplification for the \ac{MF} case has been derived in~\cite{geiger_constellation_2025}.

\begin{figure*}[!t]
    \centering
    \input{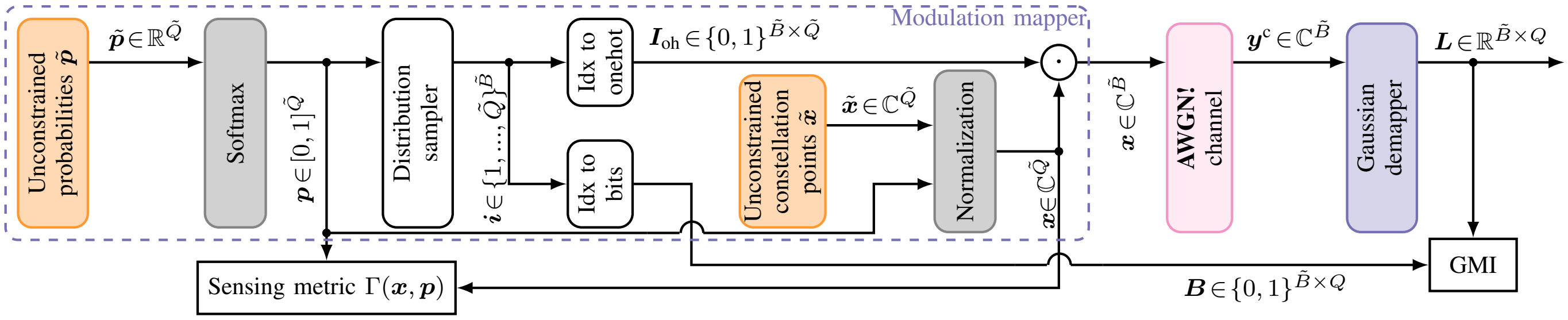}
    \vspace{-0.7cm}
    \caption{Proposed \ac{AE} framework to shape constellations for \ac{ISAC}. The trainable parameters are marked in orange and their normalization in gray.}
    \label{fig:AE:setup}
    \vspace{-0.6cm}
\end{figure*}

Once the parameters are determined from the constraint system, the corresponding differential entropy follows in closed form as
\vspace{-0.4cm}
\begin{equation}
    \mathbbm{h}(\RV{x}_{\text{max}})  = 
    \begin{cases}
        \frac{- C_{-2} \cdot \gamma^{\text{RF}}_{-2} -\gamma^{\text{RF}}_{0} - C_2 \cdot \gamma^{\text{RF}}_{2}}{\ln(2)}, & \text{RF}, \\
        \frac{-\gamma^{\text{MF}}_{0} - C_2 \cdot \gamma^{\text{MF}}_{2} - C_4 \cdot \gamma^{\text{MF}}_{4}}{\ln(2)}, & \text{MF},
    \end{cases}
    \label{eq:bounds:entropy_maximizing_distribution}
    \vspace{-0.10cm}
\end{equation}
which is obtained by substituting~\eqref{eq:Bound:Ansatz_Maximum_entropy} into the definition of the differential entropy. This allows us to compare the optimized channel input \acp{PDF} for the fundamental constraint types (see Sec.~\ref{sec:results:optimization:entropy_maximization}) at equal differential entropy, thereby providing an analytical reference for the following numerical optimization.

\vspace{-0.4cm}
\subsection{Autoencoder-based Constellation Optimization}
\vspace{-0.15cm}
\label{sec:autoencoder}
\setulcolor{DarkGreen}\RevB{The underlying differentiable bitwise \mbox{\ac{AE}} optimization framework \mbox{follows~\cite{geiger_constellation_2025}} and the references therein. The present work modifies the optimization in three key aspects: First, the \mbox{\ac{MF}}-specific kurtosis constraint is replaced by the \mbox{\ac{SCE}}-dependent sensing \mbox{metric~$\Gamma(\mathcal{X}, P_{\RV{x}}(x))$}. Second, the communications \mbox{\ac{SNR}} is sampled over a range of operating points rather than fixed during training. Third, we extend the framework from individually optimized constellations to an adaptive constellation family, as described in \mbox{Sec.~\ref{sec:opt:adaptive_constellations}}.}

\RevA{\mbox{Figure~\ref{fig:AE:setup}} shows the block diagram of the framework for \mbox{\ac{GMI}} \mbox{maximization~\eqref{eq:theory:optimization_problem:GMI_opt}}}. For a batch size $\tilde{B}$, the distribution sampler generates random indices $\vect{i} \in \{1, \ldots, \tilde{Q} \}^{\tilde{B}}$ according to the input distribution \mbox{$P_{\RV{x}}(x) \triangleq \vect{p}$}. To ease optimization, we use the Gumbel-softmax trick and optimize the unconstrained probabilities \mbox{$\tilde{\vect{p}} \in \mathbb{R}^{\tilde{Q}}$} which are passed through a softmax layer to ensure that $\vect{p}$ satisfies the probability distribution properties~(\ref{eq:theory:optimization_problem:C0})~\cite{stark_joint_2019}. These indices are then mapped to bits~$\vect{B}$ and one-hot vectors~$\vect{I}_{\mathrm{oh}}$. To select the transmit symbols, the one-hot vectors are pointwise multiplied with the normalized constellation points $\vect{x}$. Similar to the probabilities, we optimize the unconstrained constellation points $\tilde{\vect{x}}$, which are passed through a normalization layer to obtain the zero mean and unit average power constellation points \mbox{$\mathcal{X} \triangleq \vect{x}$} satisfying~\eqref{eq:system_model:zero_mean_unit_power}. The transmission of the selected constellation symbols over an \ac{AWGN} channel~\eqref{eq:system_model:comm_channel} is simulated. \RevB{In contrast to~\cite{geiger_constellation_2025}, the communications \ac{SNR} is uniformly sampled from $[\mathrm{SNR}_{\min},\mathrm{SNR}_{\max}]$ during training, such that the optimized constellation maximizes the average \ac{GMI}~\eqref{eq:theory:optimization_problem:GMI_opt} under varying effective noise powers after equalization of the \ac{LTI} communications channel.} The corresponding \acp{LLR}~(\ref{eq:system_model:LLR}) are computed for each bit using a classical Gaussian demapper~\cite{Ivanov_BICM}. Finally, the \ac{GMI}~(\ref{eq:system_model:GMI}), which should be maximized, is computed from the bits and \acp{LLR}.

To satisfy a specific sensing performance, we reformulate the sensing constraint~(\ref{eq:theory:optimization_problem:C2}) as a sensing loss that directly reflects the \ac{SCE}-specific \ac{ISINR} constraints
\vspace{-0.15cm}
\begin{equation}
    L_{\text{sens}} = {\begin{cases}
        0, &   \Gamma(\mathcal{X}, P_{\RV{x}}(x)) \leqslant \tilde{\Gamma} \\
        d (\Gamma(\mathcal{X}, P_{\RV{x}}(x)) - \tilde{\Gamma}), & \Gamma(\mathcal{X}, P_{\RV{x}}(x)) > \tilde{\Gamma},
    \end{cases}}
    \label{eq:AE:sensing_loss}
    \vspace{-0.2cm}
\end{equation}
where the penalty factor \mbox{$d \in \mathbb{R}^{+}$} controls the strength of the penalty if the sensing threshold $\tilde{\Gamma}$ is violated. The sensing metric~$\Gamma(\mathcal{X}, P_{\RV{x}}(x))$ is chosen according to the \ac{SCE} for which the constellation is optimized, as summarized in Tab.~\ref{tab:theory:Sensing_constraint}. To maximize the \ac{SC} performance, the overall loss function combines both
\vspace{-0.4cm}
\begin{equation}
    L_{\text{S\&C}} = \underbrace{\frac{Q - \mathbb{I}_{\text{GMI}}}{Q}}_{\text{Communications}} + L_{\text{sens}}.
    \label{eq:AE:overall_loss}
    \vspace{-0.1cm}
\end{equation}

Finally, depending on geometric, probabilistic or joint shaping, we optimize the corresponding parameters, i.e., constellation points, their probabilities or both, by minimizing the loss function~(\ref{eq:AE:overall_loss}) using gradient descent. %

\vspace{-0.55cm}
\subsection{Adaptive Constellations}\label{sec:opt:adaptive_constellations}
\vspace{-0.15cm}
\setulcolor{blue}\RevC{In practice, the constellation must be adapted to the current \mbox{\ac{SC}} requirements and sensing conditions. Therefore, we extend the threshold-specific optimization from \mbox{Sec.~\ref{sec:autoencoder}} to adaptive constellation \mbox{shaping~\cite{cammerer_trainable_2020}} for \mbox{\ac{ISAC}}. Specifically, we optimize a continuous constellation \mbox{family~$\boldsymbol{\zeta} \mapsto \left(\mathcal{X}(\boldsymbol{\zeta}),P_{\RV{x}}(\cdot;\boldsymbol{\zeta})\right)$}, where \mbox{$\boldsymbol{\zeta}=\tilde{\Gamma}$} for the \mbox{\ac{MF}} and \mbox{\ac{RF}}, and \mbox{$\boldsymbol{\zeta}=(\tilde{\Gamma},\tilde{\sigma}_{\mathrm{s}}^2)$} for the \mbox{\ac{WF}}. To this end, depending on the shaping method, we replace the input-independent trainable constellation points, the probability logits, or both by \mbox{\acfp{NN}} that \mbox{take~$\boldsymbol{\zeta}$} as input. Their outputs are passed through the same normalization and softmax operations as in \mbox{Sec.~\ref{sec:autoencoder}}, ensuring zero mean, unit average power constellations and valid probability distributions. During training, the \mbox{\acp{NN}} are jointly optimized by sampling a batch of operating \mbox{points~$\boldsymbol{\zeta}$}, generating the corresponding constellations, and averaging the overall \mbox{\ac{SC}} \mbox{loss~\eqref{eq:AE:overall_loss}} over the batch. After training, the \mbox{\acp{NN}} can generate the constellation points and probabilities for a desired operating \mbox{point~$\boldsymbol{\zeta}$} by a single forward pass. Here, the sensing \mbox{threshold~$\tilde{\Gamma}$}, i.e., \mbox{\ac{ISINR}}, could be obtained from a required detection probability by \mbox{inverting~\eqref{eq:pd_isinr}}, while, for the \mbox{\ac{WF},~$\tilde{\sigma}_{\mathrm{s}}^2$} can be estimated from the preceding transmission.}

\vspace{0.3cm}
\section{Experimental and Simulation Setup}
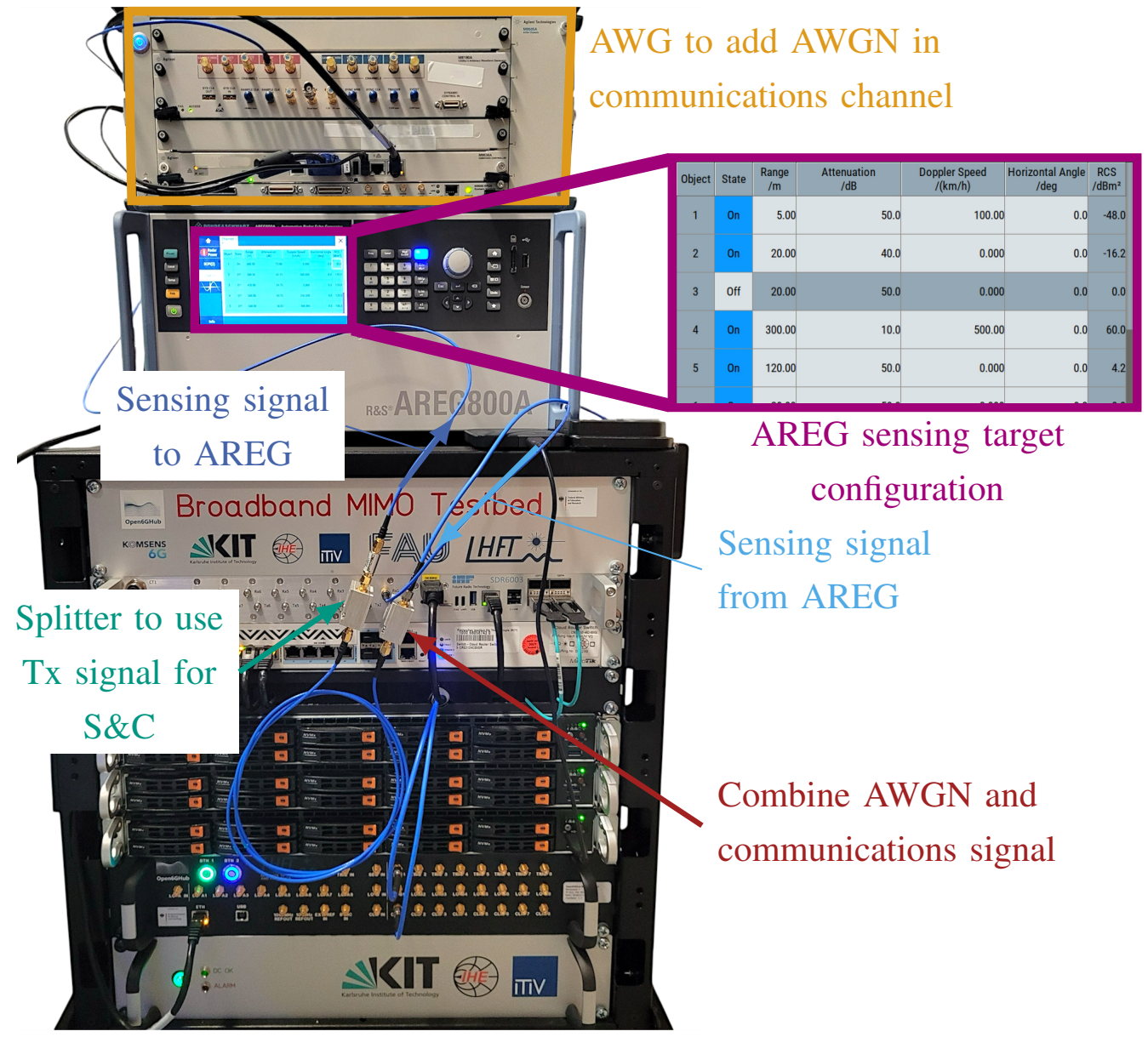
\begin{figure}[!t]
    \centering
    \begin{tikzpicture}[>=latex]
        \node[anchor=south west, inner sep=0] (img) at (0,0)
        {\includegraphics[width=0.6\linewidth]{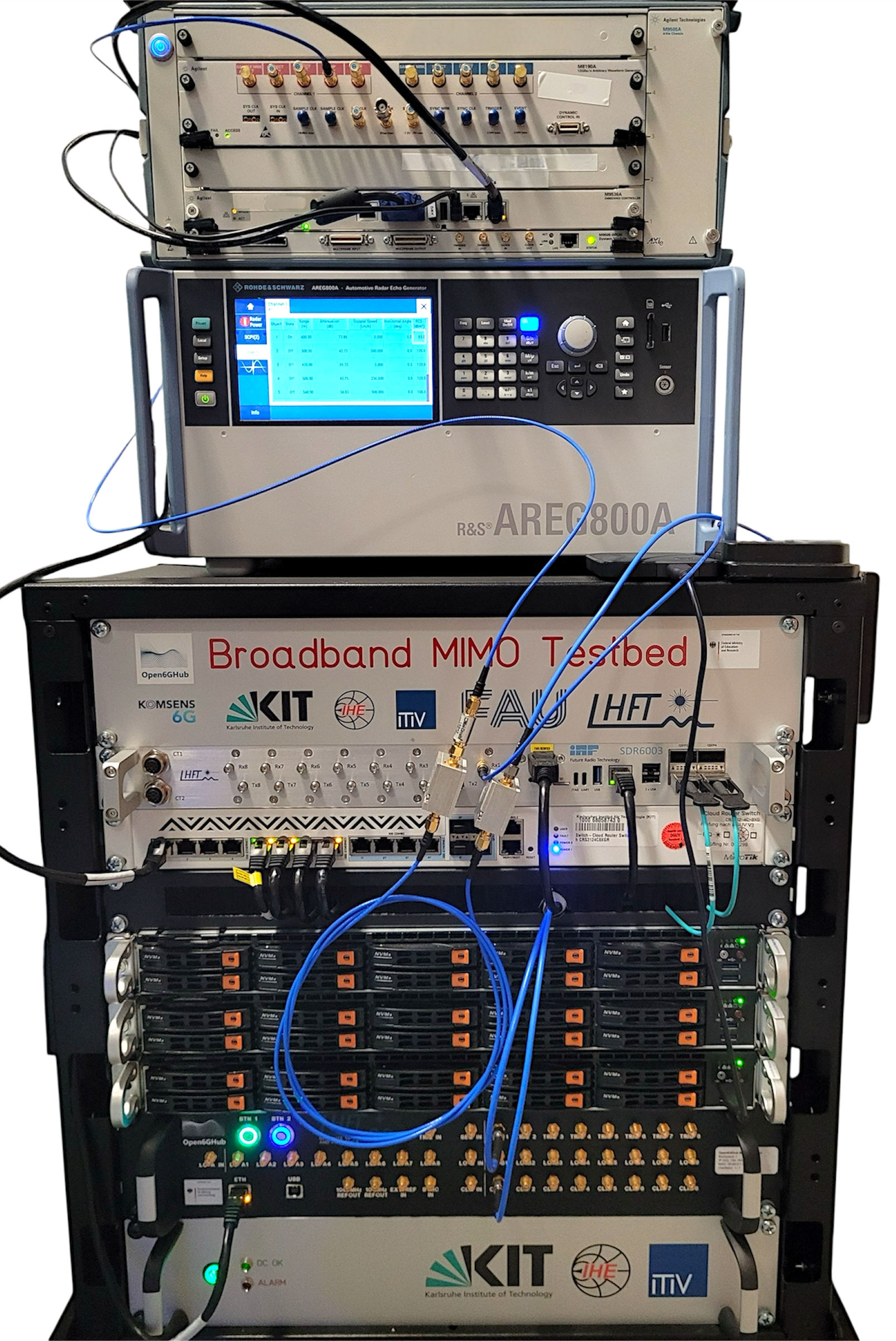}};
        \begin{scope}[x={(img.south east)}, y={(img.north west)}]
            \draw [line width = 2pt, KITorange] (0.163,0.995) -- (0.81,0.995) -- (0.8,0.81) node [pos=0.3, anchor=west, align=left] {\footnotesize AWG to add AWGN in\\ \footnotesize communications channel} -- (0.168,0.81) -- (0.163,0.999);
            \draw [line width = 2pt, KITpurple] (0.26,0.78) -- (0.49,0.78) -- (0.49,0.685) -- (0.26, 0.685) -- (0.26,0.785);
            \node[anchor=west, inner sep=0, draw, line width=4pt, KITpurple] (img2) at (0.95,0.73)
            {\includegraphics[width=0.4\linewidth]{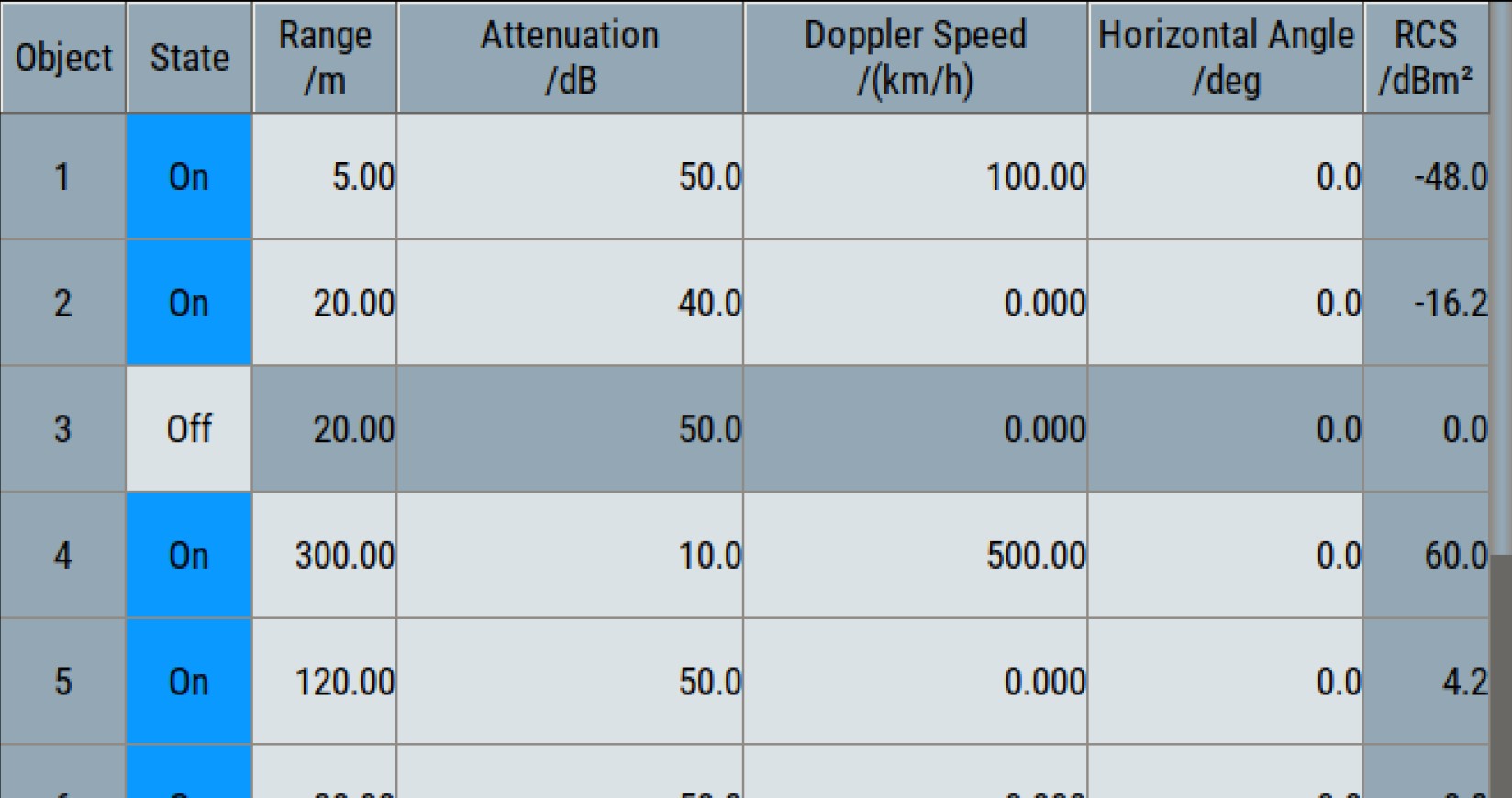}};
            \draw [line width = 3pt, KITpurple] (0.49,0.78) -- (0.953,0.85);
            \draw [line width = 3pt, KITpurple] (0.49,0.685) -- (0.953,0.61);
            \node [KITpurple, anchor=north, align=center] at (1.3,0.61) {\footnotesize AREG sensing target\\ \footnotesize configuration};
            \node[fill=white, text = KITgreen, align=center] at (0.15,0.35) {\footnotesize Splitter to use\\ \footnotesize Tx signal for\\ \footnotesize S\&C};
            \draw [->, very thick, KITgreen] (0.325, 0.35) -- (0.48, 0.42);
            \draw [KITblue, ->, very thick] (0.56,0.53) -- (0.64,0.6);
            \node [anchor=east, align=center, fill=white, text=KITblue] at (0.48,0.59){\footnotesize Sensing signal\\ \footnotesize to AREG};
            \draw [-, KITblue] (0.60,0.565) -- (0.48, 0.59);
            \node [anchor=west, text=KITred, align=left] at (1.0, 0.2) {\footnotesize Combine AWGN and\\ \footnotesize communications signal};
            \draw [->, very thick, KITred] (1.0, 0.2) -- (0.57, 0.39);
            \node [anchor=west, text=KITcyanblue, align=left] at (1.0, 0.45) {\footnotesize Sensing signal\\ \footnotesize from AREG};
            \draw [->, very thick, KITcyanblue] (0.76, 0.57) -- (0.61, 0.47);
            \draw [-, KITcyanblue] (0.685, 0.52) -- (1.0,0.45);
        \end{scope}
    \end{tikzpicture}
    \vspace{-0.65cm}
    \caption{Experimental setup of the \ac{OFDM}-based \ac{ISAC} system.
    The transmit signal generated by the Broadband Massive MIMO testbed is physically split into sensing and communications channels. The sensing channel is emulated using a Rohde \& Schwarz AREG, while the communications channel is impaired by additive noise with controlled power.}
    \label{fig:setup:experimental_setup}
    \vspace{-0.65cm}
\end{figure}

\vspace{-0.1cm}
\RevB{To verify the theoretical analysis, we experimentally investigate an \ac{OFDM}-\ac{ISAC} system with a \ac{5G} New Radio numerology \mbox{$\mu=\num{2}$} in a hardware-in-the-loop setup rather than in an over-the-air measurement.} The experiments are conducted using the Broadband Massive MIMO testbed~\cite{Nuss_Testbed}, shown in Fig.~\ref{fig:setup:experimental_setup}. The testbed comprises \num{32} baseband modules based on radio-frequency system-on-chips and \num{64} backend servers, enabling scalable and fully software-defined operation.

\RevA{To ensure a fair comparison, all waveforms are normalized to the same average transmit power and clipped at \SI{12}{dB} above it. As \(N\) increases, each sample of the \ac{OFDM} signal in the time domain approaches a complex Gaussian distribution due to the central limit theorem. Consequently, constellation shaping mainly affects the \ac{PAPR} for small \(N\), whereas its influence diminishes for large \(N\)~\cite{ochiai_distribution_2001}.} To enable \ac{ISAC}, the transmit signal generated by the testbed is physically split into a communications and a sensing path, following Fig.~\ref{fig:system_model:block_diagram}. To emulate the sensing channel, the transmit signal is passed through a Rohde \& Schwarz automotive radar echo generator (AREG) 800A, \RevB{which can emulate up to \num{8} targets, each characterized by an individual delay, Doppler shift, and \ac{RCS} (see the zoom-in in Fig.~\ref{fig:setup:experimental_setup})~\cite{areg_datasheet}.} \setulcolor{red}\RevC{Radar echo generators are established industrial tools for the development and validation of automotive radar sensors and radar-based advanced driver assistance systems functions, as they enable controlled and reproducible testing conditions\mbox{~\cite{diewald_radar_2021,areg_datasheet}}.} \setulcolor{blue}\RevC{The controlled setup was deliberately chosen to enable a quantitative validation of the proposed analytical \mbox{\ac{ISINR}} framework.} \RevB{The power of the \ac{TOI} or the interfering target is varied over \SI{70}{dB}, spanning a broad range of practically relevant target echo strengths, e.g., from a nearby vehicle with large \ac{RCS} to a distant target with a comparatively small \ac{RCS} such as a drone.} In the communications path, noise generated by an \ac{AWG} is added to the transmit signal, enabling controlled variation of the communications \ac{SNR}. The received sensing and communications signals are sampled on separate channels and processed independently. All baseband signal processing is performed offline in MATLAB, and all experimental results are averaged over \num{60} measurements.

\begin{table}[!b]
    \vspace{-0.65cm}
    \caption{Considered system parametrizations}
    \vspace{-0.25cm}
    \label{tab:ch5:system_parameter}
    \centering
    \begin{tabular}{cccc}
        \toprule
        Parameter & Symbol & Full frame & Partial frame \\
        \midrule
        Effective bandwidth & $B$ & \num{800} MHz & \num{100} MHz  \\
        No. used subcarriers & $N$ & \num{13334} & \num{1666} \\
        No. \ac{OFDM} symbols & $M$ & \num{560} & \num{32} \\
        Processing gain & $G_{\text{P}}$ & \SI{66.04}{dB} & \SI{44.48}{dB} \\
        \bottomrule
    \end{tabular}
\end{table}

The experimental setup operates at a carrier frequency of \mbox{$f_{\mathrm{c}}=\SI{3.6864}{GHz}$} with a sampling rate of \SI{983.04}{MSa/s}, resulting in \mbox{$N_{\text{total}}=\num{16384}$} subcarriers and a subcarrier spacing of \mbox{$\Delta f=\SI{60}{kHz}$}. Only the $N$ central subcarriers are utilized, and by varying their number, the effective bandwidth is controlled. Two configurations are considered and summarized in Tab.~\ref{tab:ch5:system_parameter}: a full frame configuration with an effective bandwidth of \SI{800}{MHz} and \mbox{$M=\num{560}$} \ac{OFDM} symbols, corresponding to a \SI{10}{ms} radio frame, and a partial frame configuration with an effective bandwidth of \SI{100}{MHz} and \mbox{$M=\num{32}$} \ac{OFDM} symbols, representative of typical \ac{5G} transmission. The cyclic prefix length is set to \num{7}\% of the useful \ac{OFDM} symbol duration. The sensing receiver applies a Hamming window prior to range-Doppler processing, using oversampling factors of \mbox{$\eta_\tau = \eta_\text{D} = 16$}. All simulations follow the experimental setup and are configured using the same system parameters. %

\vspace{-0.1cm}
\section{Sensing Performance Under Different \acp{SCE}: Experimental and Simulation Results}
\vspace{-0.1cm}
\label{sec:results:SCE}
This section verifies the derived impact of the \acp{SCE} and conventional modulation formats on the sensing performance through simulations and experiments.

\vspace{-0.3cm}
\subsection{Single Target Scenario}
\begin{figure}[!t]
    \centering
    \input{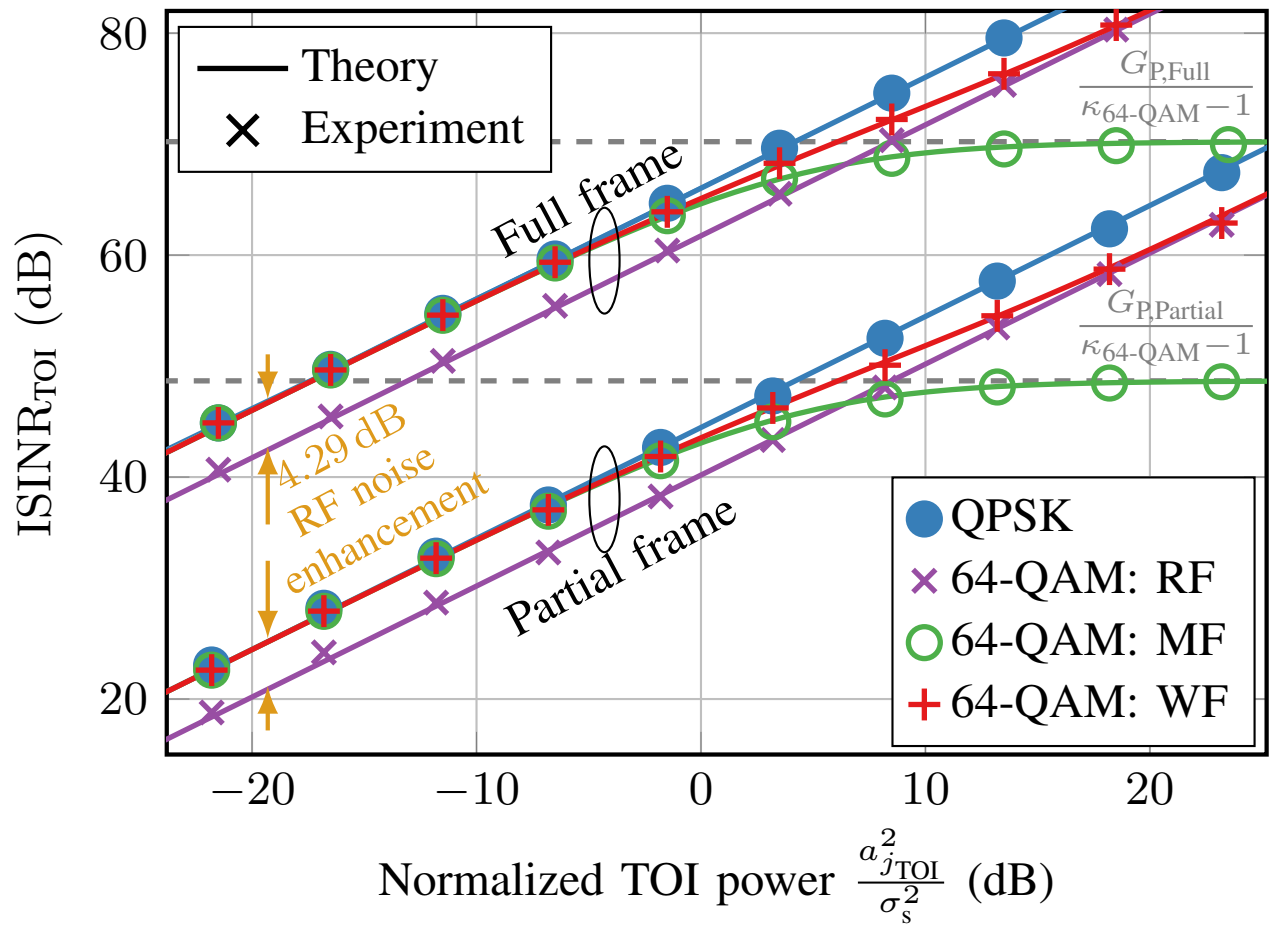}
    \vspace{-0.4cm}
    \caption{\ac{ISINR} versus normalized target power for a single target scenario using different \acp{SCE}, comparing \ac{QPSK} and \num{64}-\ac{QAM}. Results are shown for full frame and partial frame parameterizations. Markers denote experimental measurements, while solid lines show the derived \ac{ISINR} from Tab.~\ref{tab:ch3:Image_SINR}.}
    \label{fig:results:sensing_channel_estimator_comparison:single_target}
    \vspace{-0.2cm}
\end{figure}

\vspace{-0.1cm}
We first consider a single target scenario and compare the resulting \ac{ISINR} for \ac{QPSK} and \num{64}-\ac{QAM} across the considered \acp{SCE}. For both parameterizations, we sweep the power of the target, and measure the resulting \ac{ISINR}. The results are shown in Fig.~\ref{fig:results:sensing_channel_estimator_comparison:single_target}, where markers denote the experimental results and solid lines represent the derived \ac{ISINR} from Sec.~\ref{sec:theory:SCE}. The close agreement between theory and experiment confirms our analytical derivations summarized in Tab.~\ref{tab:ch3:Image_SINR}. Our simulation results show the same performance and are omitted for visual clarity.

For \ac{QPSK}, all three \acp{SCE} coincide (up to scaling), resulting in the same \ac{ISINR}, which increases linearly with the normalized target power for both parameterizations. The full frame configuration outperforms the partial frame by \SI{21.56}{dB}, consistent with its higher coherent processing gain~(\ref{eq:theory:SCE:processing_gain}). \RevA{The linear increase of the measured \ac{ISINR} with the target power, without observable compression, indicates that the experimental setup operates in its linear regime. Residual nonlinear distortion can therefore be considered negligible in the measurements.}

For \num{64}-\ac{QAM}, the \ac{ISINR} is generally equal to or lower than that of \ac{QPSK}, depending on the normalized target power and the chosen \ac{SCE}. The \ac{RF} introduces noise enhancement, resulting in an \ac{ISINR} reduction of \SI{4.29}{dB} compared to \ac{QPSK}, independent of the normalized target power, in agreement with the predicted noise enhancement~(\ref{eq:theory:SCE:RF:noise_enhancement}). In contrast, the \ac{MF} introduces interference proportional to the kurtosis $\kappa$ of the constellation and the target power. As a result, the \num{64}-\ac{QAM} achieves the same performance as \ac{QPSK} in the low-\ac{SNR} regime, where the \ac{ISINR} is dominated by the noise floor, but saturates at higher input \ac{SNR} at the predicted \ac{ISINR}~(\ref{eq:theory:SCE:MF:Image_SINR_saturation}).

As expected, the \ac{WF} combines the benefits of both \acp{SCE}: it coincides with the \ac{MF} at low \ac{SNR} and approaches the \ac{RF} performance at high \ac{SNR}, outperforming both in the intermediate regime. Consequently, while the \ac{MF} is favorable at low \ac{SNR} and the \ac{RF} at high \ac{SNR}, the \ac{WF} consistently achieves the highest \ac{ISINR} among the considered \acp{SCE} by optimally balancing noise enhancement and interference.

\begin{figure}[!t]
    \centering
    \input{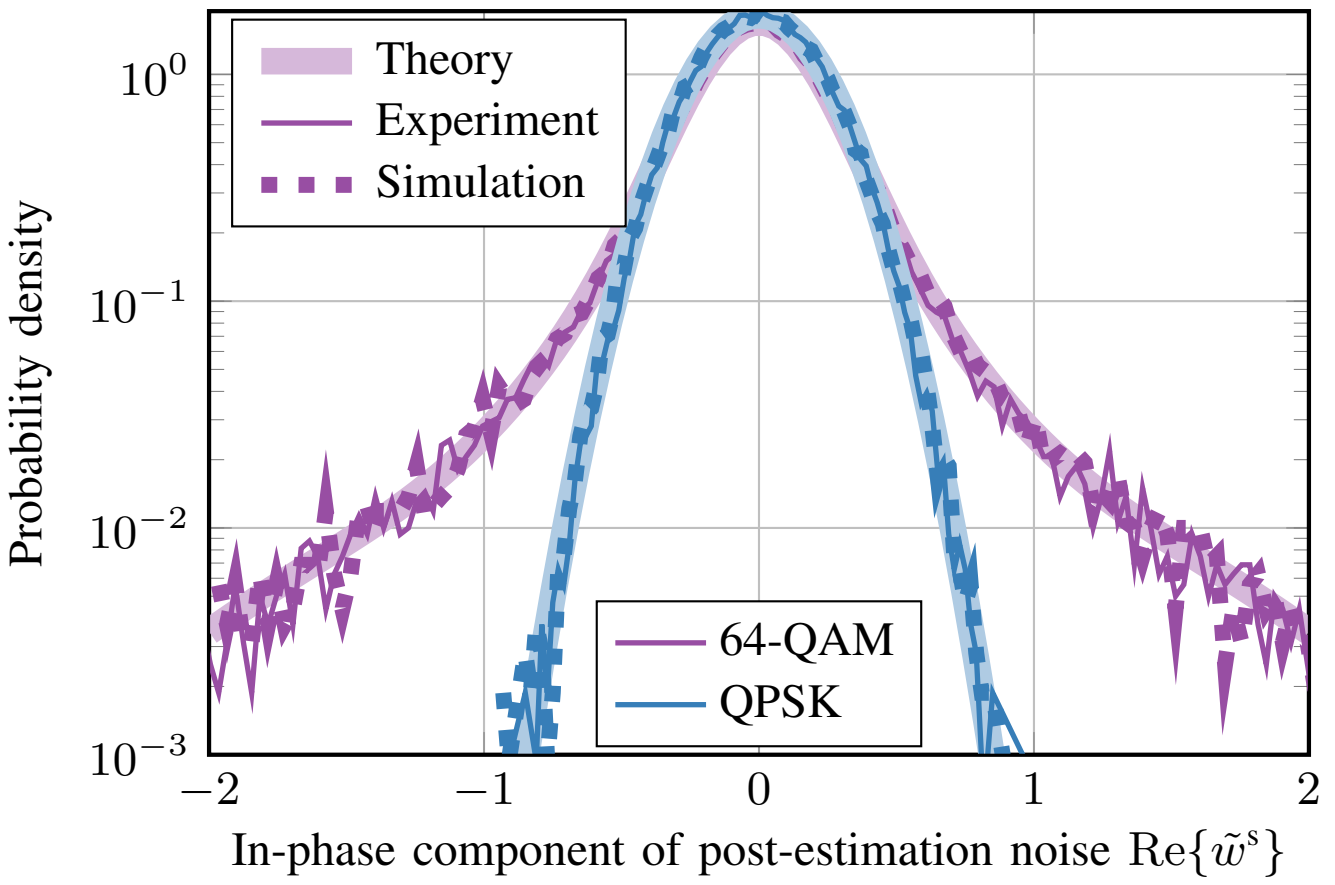}
    \vspace{-0.8cm}
    \caption{In-phase component of the theoretical, simulated, and experimental noise \ac{PDF} after \ac{RF} processing for \ac{QPSK} and \num{64}-\ac{QAM} at a normalized target power of \SI{10}{dB}.}
    \label{fig:results:noise_distribution_SD}
    \vspace{-0.2cm}
\end{figure}

Fig.~\ref{fig:results:noise_distribution_SD} shows the in-phase component of the noise distribution after the \ac{RF} for \ac{QPSK} and \num{64}-\ac{QAM} at a normalized target power of \SI{10}{dB}. Theoretical, experimental, and simulated \acp{PDF} are shown for comparison. The results confirm our analytical predictions: for \ac{QPSK}, the noise follows a Gaussian distribution (parabolic in the logarithmic domain), whereas for \num{64}-\ac{QAM}, pronounced tails emerge. These tails originate from small amplitude constellation points, which induce noise amplification through the division operation in the \ac{RF}, as predicted by the analysis in Sec.~\ref{sec:theory:SCE:mean_var:RF}. While detection performance depends on the noise-and-interference power, range and velocity estimation are influenced by the post-estimation noise \ac{PDF}. Consequently, future work should investigate the impact of non-Gaussian distributed noise on the estimation performance.

\vspace{-0.3cm}
\subsection{Multi Target Scenario}
\begin{figure}[!t]
    \centering
    \input{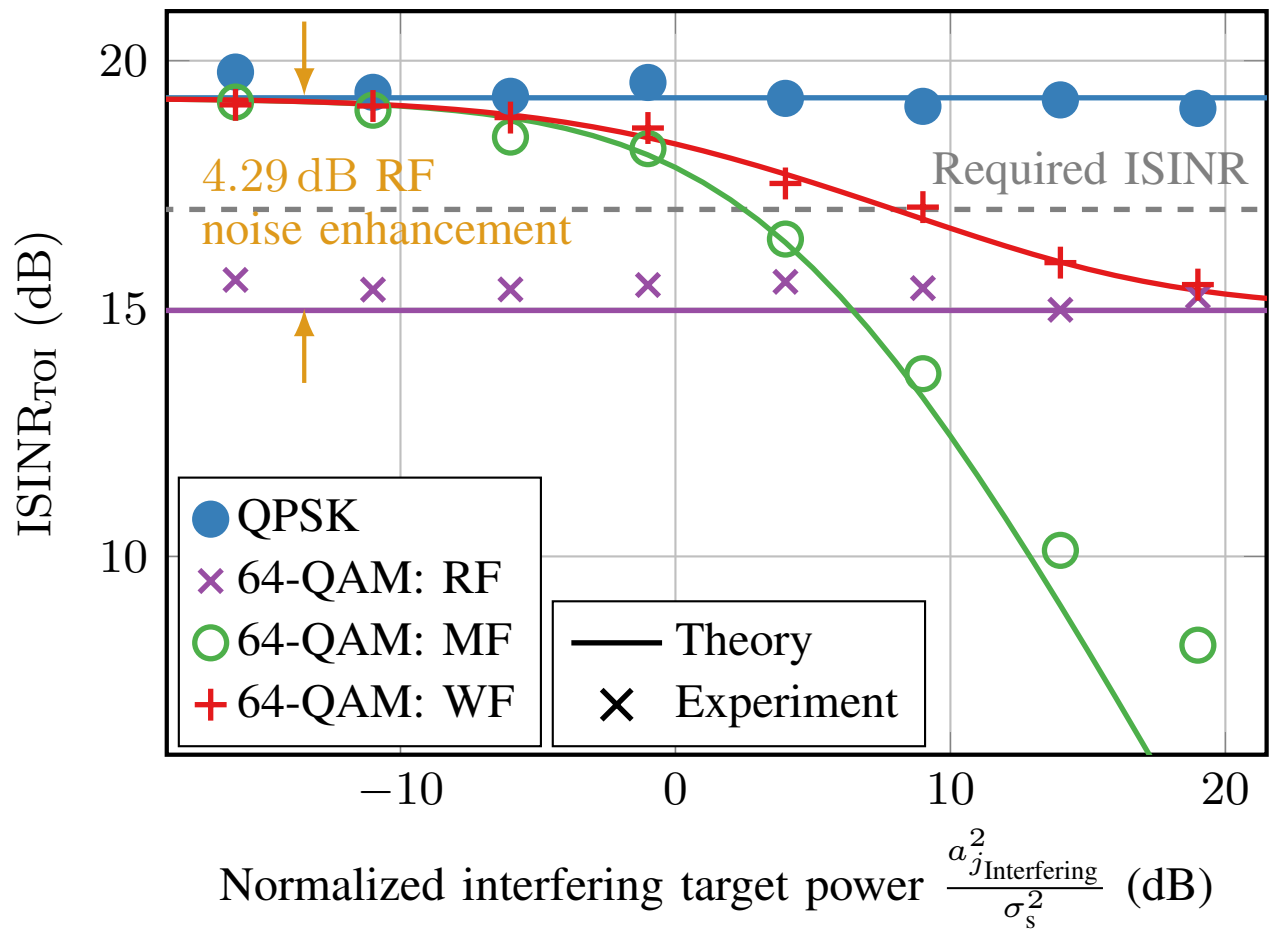}
    \vspace{-0.2cm}
    \caption{\ac{ISINR} of the target of interest as a function of the power of the interfering target in a two target scenario. Results are shown for a small frame configuration, comparing different \acp{SCE} for \ac{QPSK} and \num{64}-\ac{QAM}. Markers denote experimental measurements, while solid lines show the derived \ac{ISINR} from Tab.~\ref{tab:ch3:Image_SINR}.}
    \label{fig:results:sensing_channel_estimator_comparison:two_targets}
    \vspace{-0.2cm}
\end{figure}

\vspace{-0.1cm}
We extend the analysis to a multi (two) target scenario consisting of a \ac{TOI} with fixed power and an interfering target whose power is swept. The \ac{ISINR} of the \ac{TOI} is evaluated as a function of the power of the interfering target for a small frame configuration. \RevB{We emulate the practically relevant and particularly challenging scenario of detecting a distant drone in the presence of substantially stronger reflectors. Accordingly, the \ac{TOI} is configured as a target with a small \ac{RCS}.} Fig.~\ref{fig:results:sensing_channel_estimator_comparison:two_targets} compares the resulting \ac{ISINR} for different \acp{SCE} using \ac{QPSK} and \num{64}-\ac{QAM}. In all cases, the experimental results validate the derived \ac{ISINR}. 

For \ac{QPSK} and for \num{64}-\ac{QAM} with \ac{RF}, the \ac{ISINR} is independent of the power of the interferer, with the latter exhibiting a lower \ac{ISINR} due to noise enhancement. In contrast, for \num{64}-\ac{QAM} with \ac{MF}, the \ac{ISINR} of the \ac{TOI} decreases with increasing interferer power due to interference that scales with the kurtosis $\kappa$ of the constellation and the accumulated power of the targets. The \ac{WF} outperforms both by balancing interference and noise enhancement. 

\begin{table}[!b]
    \centering
    \caption{Measured and theoretically predicted \ac{ISINR} in the presence of clutter (cl). All values are given in dB.}
    \vspace{-0.25cm}
    \label{tab:results:clutter}
    \begin{tabular}{ccccccc}
        \toprule
        & RF & & \raisebox{-3mm}{\makecell{Target \\ No.}} & & MF & \\[-5mm] \cmidrule(lr){1-3}\cmidrule(lr){5-7}
        \makecell{Theory\\w/o cl} & \makecell{Theory\\w/ cl} & Measured & & Measured & \makecell{Theory\\w/ cl} & \makecell{Theory\\w/o cl} \\
        \midrule
        55.04 & 55.04 & 54.75 & 1/2 & 40.96 & 39.76 & 44.42 \\
        50.04 & 50.04 & 49.96 & 3/4 & 34.67 & 34.76 & 39.42 \\
        35.96 & 35.96 & 36.41 & 5/6 & 21.33 & 20.67 & 25.34 \\
        \bottomrule
    \end{tabular}
\end{table}

Assuming a required \ac{ISINR} of \SI{17}{dB} for reliable detection~\cite{mandelli_survey_2023}, the results show that the \ac{TOI} is not reliably detectable in the presence of a strong interfering target when conventional higher-order modulation formats are used. This highlights the fundamental challenge of jointly achieving high throughput and reliable sensing, motivating the need for optimized constellations.

\vspace{-0.6cm}
\setulcolor{DarkGreen}
\subsection{Clutter}
\RevA{To assess the impact of unexpected clutter, we consider a scenario with eight reflections, of which two zero-Doppler reflections are interpreted as clutter and the six others are treated successively as \mbox{\ac{TOI}}. \mbox{Tab.~\ref{tab:results:clutter}} compares the measured and theoretically predicted \mbox{\ac{ISINR}} of the six \mbox{\acp{TOI}} for known and unknown clutter. When the clutter is included in the model, the measured \mbox{\ac{ISINR}} closely follows the theoretical prediction. For the \mbox{\ac{RF}}, the predicted \mbox{\ac{ISINR}} is identical with and without clutter, since the \mbox{\ac{RF}} introduces only noise enhancement, which is independent of additional reflections. In contrast, for the \mbox{\ac{MF}}, neglecting the clutter leads to an overestimation of the \mbox{\ac{ISINR}}, as the additional reflections caused by clutter increase the signaling-induced interference. This experiment therefore confirms that clutter must be accounted for in the model and that the proposed analysis remains valid when such components are included.}

\vspace{0.3cm}
\section{ISAC Performance under Constellation Shaping: Experimental and Simulation Results}
\vspace{-0.1cm}
In this section, we present the entropy-maximizing channel input \acp{PDF}, optimize constellations for \ac{ISAC} using the unified framework, and quantify and compare the resulting performance gains. Throughout this section, we quantify the sensing performance through the \ac{ISINR} in a two target scenario (one strong interferer and a weak \ac{TOI}). The average communications \ac{SNR} $1/\tilde{\sigma}^2_{\text{c}}$ after equalization is \SI{10.15}{dB}, where \mbox{$\tilde{\sigma}^2_{\text{c}} = \frac{1}{N}\sum_{n=0}^{N-1}\sigma^2_{\text{c,}n}$}. %

\begin{figure}[!t]
    \centering
    \input{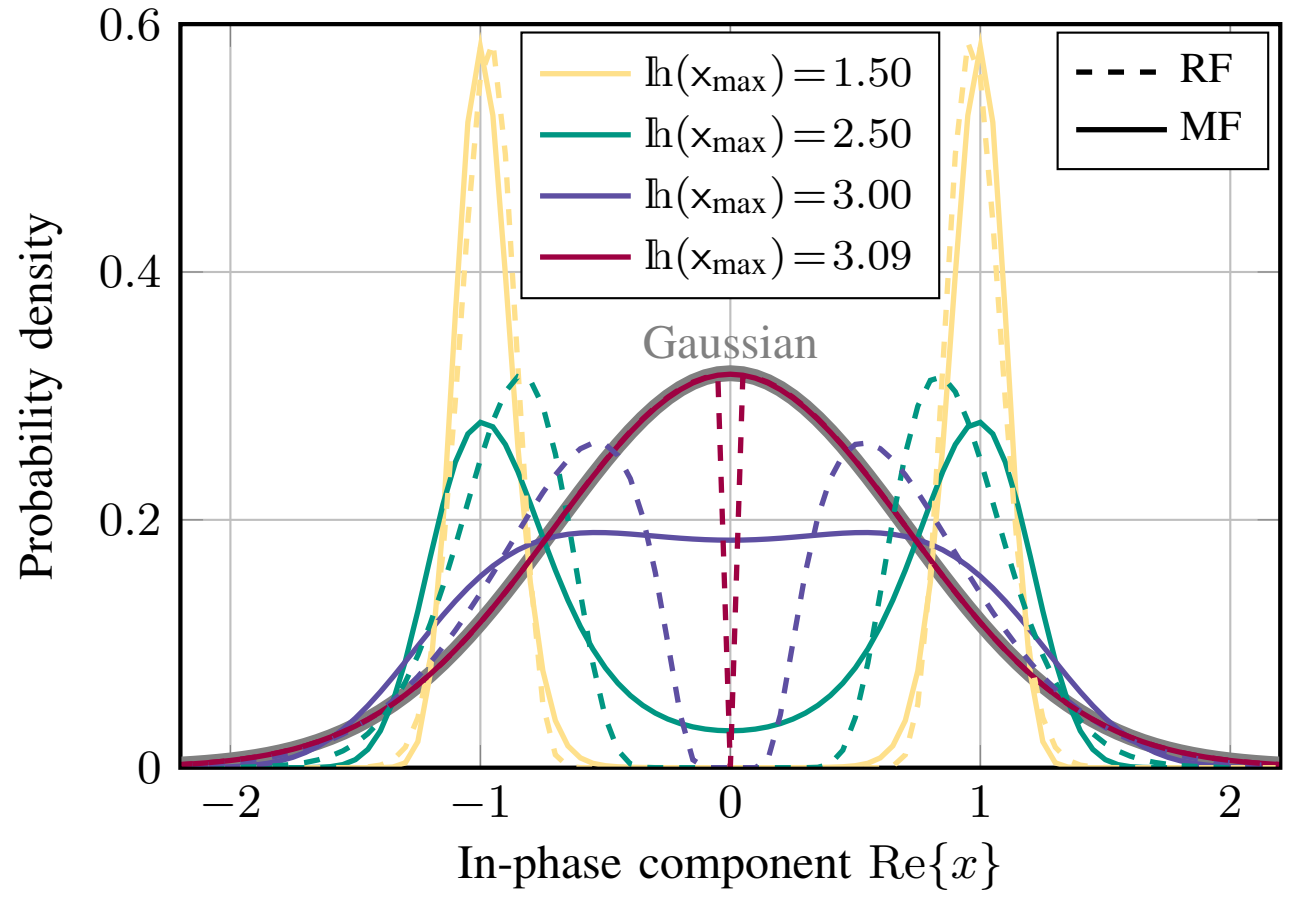}
    \vspace{-0.4cm}
    \caption{One-dimensional radial cuts of the entropy-maximizing input \acp{PDF} for the \ac{MF} and \ac{RF} cases at different differential entropy values. Due to circular symmetry, the angular direction is irrelevant.}
    \label{fig:entropy_maximization:distribution_comparison}
    \vspace{-0.4cm}
\end{figure}

\begin{figure}[!t]
            \centering
            \input{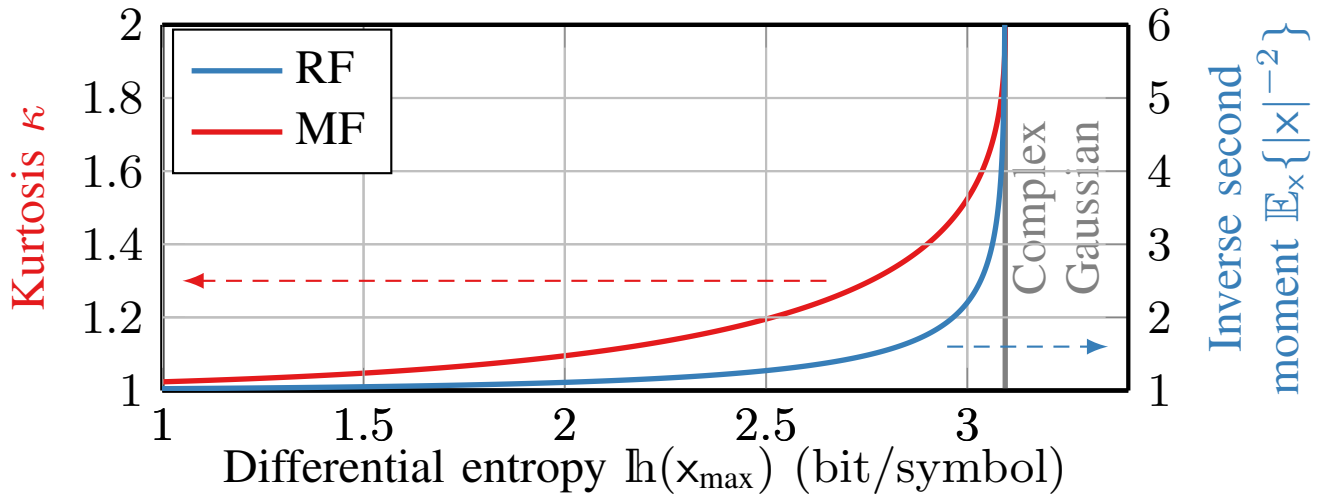}
            \vspace{-0.5cm}
            \caption{Relationship between the differential entropy of the entropy-maximizing input distribution and the sensing metric~$\Gamma(\mathcal{X}, P_{\RV{x}}(x))$, i.e., the kurtosis for the \ac{MF} and inverse second moment for the \ac{RF}.}
            \label{fig:entropy_maximization:moment_comparison}
    \vspace{-0.5cm}
\end{figure}

\vspace{-0.35cm}
\subsection{Entropy-Maximizing Channel Input \acp{PDF}}
\vspace{-0.1cm}
\label{sec:results:optimization:entropy_maximization}
Fig.~\ref{fig:entropy_maximization:distribution_comparison} shows a slice of the entropy-maximizing input \acp{PDF} for various differential entropies $\mathbbm{h}(\RV{x}_{\text{max}})$ at fixed unit average input power, which serves as an indicator for the resulting system performance. We can observe that the entropy-maximizing \acp{PDF} are similar for the extreme points of the \ac{SC} trade-off. For strong sensing constraints and consequently low entropy, e.g., \mbox{$\mathbbm{h}(\RV{x}_{\text{max}}) = 1.5$}, the \acp{PDF} approach a constant-modulus distribution, which is sensing-optimal for all considered \acp{SCE}. For an entropy close to that of a communications-optimal circularly symmetric complex Gaussian, i.e., \mbox{$\mathbbm{h}(\RV{x}_{\text{max}})=3.09$}, the entropy-maximizing input \acp{PDF} closely resemble a Gaussian distribution for both the \ac{MF} and the \ac{RF}. In the \ac{RF} case, a pronounced dip around the origin enforces the convergence of the inverse second moment $\expecv{\RV{x}}{|\RV{x}|^{-2}}$, whereas for the \ac{MF}, the Gaussian optimum is attained at a kurtosis of \mbox{$\kappa=2$}. Consequently, the entropy of a circularly symmetric complex Gaussian can be achieved exactly for the \ac{MF}, but only asymptotically for the \ac{RF}. \RevC{This distinct behavior is further illustrated in Fig.~\ref{fig:entropy_maximization:moment_comparison}, which relates the achievable differential entropy directly to the corresponding sensing-induced moment constraint.} At intermediate entropy values, the entropy-maximizing \acp{PDF} differ due to the distinct amplitude penalties imposed by each \ac{SCE}. The \ac{RF} suppresses small amplitudes to limit noise enhancement, resulting in a clear dip around the origin, while the \ac{MF} reduces large amplitude excursions by concentrating probability mass around the unit circle. Thus, the \ac{MF} and \ac{RF} impose complementary constraints on the amplitude distribution, both leading to near-constant-modulus signaling under strict sensing constraints.

\begin{figure}[!t]
    \vspace{-0.9cm}
    \input{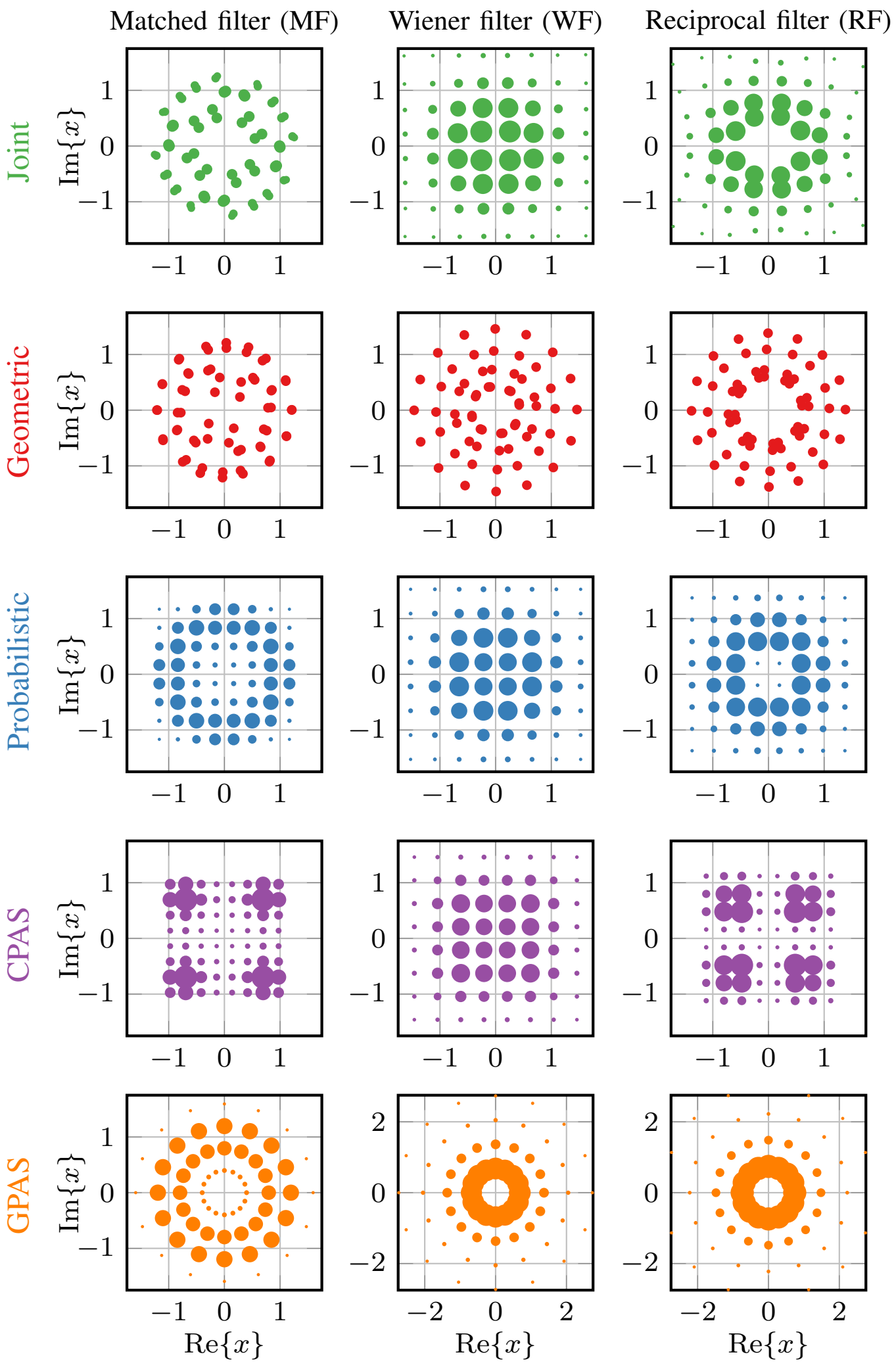}
    \vspace{-0.9cm}
    \caption{Optimized constellations for geometric, probabilistic, and joint constellation shaping as well as \ac{CPAS} and \ac{GPAS} for the considered \acp{SCE} under an intermediate sensing metric threshold \mbox{$\tilde{\Gamma} = \num{1.6}$}. The size of each constellation point is proportional to its probability and each constellation point has an associated bit label, which is omitted for clarity.}
    \label{fig:results:constellation_comparison}
\end{figure}

\vspace{-0.3cm}
\subsection{Autoencoder-based Constellation Optimization}
\vspace{-0.1cm}
All optimized constellations have \mbox{$\tilde{Q} = \num{64}$} constellation points, and the respective sensing thresholds $\tilde{\Gamma}$ are swept between the sensing- and communications-optimal regimes (see Tab.~\ref{tab:theory:Sensing_constraint}). For each \ac{SCE} and threshold, the constellations are optimized independently. For the \ac{WF}, the normalized noise power is set to \mbox{$\tilde{\sigma}^2_{\text{s}} = 0.5$}.

The constellations are optimized by minimizing the loss function~(\ref{eq:AE:overall_loss}) using the Adam optimizer. During training, the communications \ac{SNR} is uniformly varied between \mbox{$\mathrm{SNR}_{\text{min}} = \SI{8}{dB}$} and \mbox{$\mathrm{SNR}_{\text{max}} = \SI{13}{dB}$}, while an increasing penalty factor $d$ enforces the desired sensing constraint. The batch size is progressively increased from $500$ to $10{,}000$, while the learning rate is gradually decreased. The initial learning rate depends on the shaping method and the selected sensing estimator.

\subsubsection{Optimized Constellations}
\label{results:subsec:optimized-constellations}
Fig.~\ref{fig:results:constellation_comparison} shows the optimized constellations for geometric, probabilistic (generic, \ac{CPAS}, and \ac{GPAS}~\cite{geiger_constellation_2025}), and joint shaping under an intermediate threshold \mbox{$\tilde{\Gamma} = 1.6$}. To facilitate a low-complexity implementation without error propagation in the forward-error correction, \ac{CPAS} and \ac{GPAS} factorize into real and imaginary components as well as into a radial distribution with uniform phase, respectively~\cite{geiger_constellation_2025, bocherer_bandwidth_2015}.

Consistent with the entropy-maximization results in Sec.~\ref{sec:results:optimization:entropy_maximization}, we observe that constellations optimized for the \ac{RF} avoid small amplitude symbols across all shaping methods, while those optimized for the \ac{MF} exhibit reduced amplitude variations. The \ac{WF} strikes a balance between these two constraints, allowing both small amplitude symbols and reduced amplitude variations to some extent and resulting in a distribution that already resembles a circularly symmetric Gaussian. \RevC{While geometric shaping yields ring-like constellations favorable for sensing, specific bit levels collapse to improve the reliability of others. Although this increases the uncoded bit error rate, practical systems employ forward error correction, which compensates for these errors. Ultimately, the improved reliability in the remaining bit levels outweighs the increased uncoded bit error rate, maximizing the overall \ac{GMI}.} In contrast, probabilistic shaping retains a regular \ac{QAM} lattice. For \ac{CPAS}, the number of used constellation points is reduced as a result of the factorized probability of occurrence, yielding \ac{QPSK}-like constellations. \ac{GPAS} produces radially organized constellations with non-uniform symbol probabilities, thereby bridging geometric and probabilistic shaping. Joint shaping exhibits both geometric and probabilistic features.

\begin{figure}[!t]
            \centering
            \input{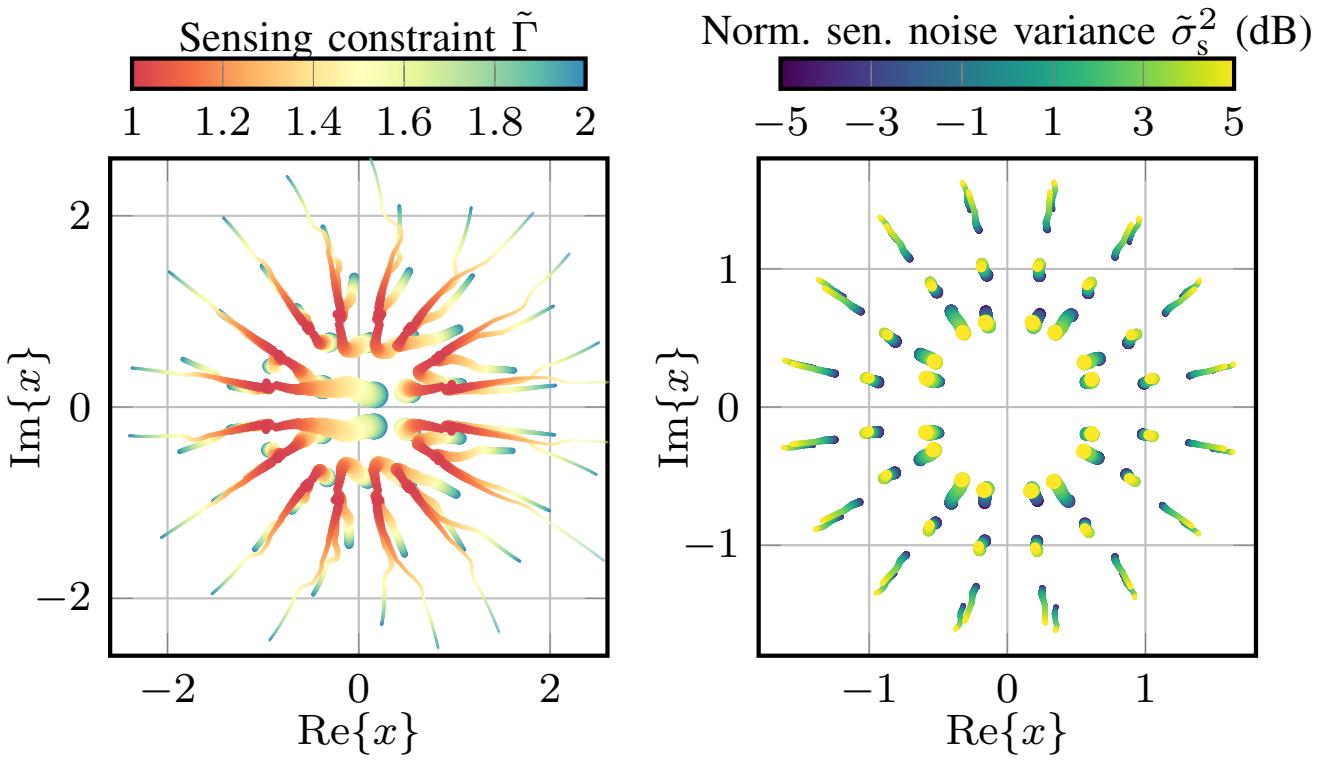}
            \vspace{-0.45cm}
            \caption{Optimized adaptive constellation for the \ac{WF} with geometric and probabilistic shaping. The left panel shows the constellation as a function of the sensing constraint~$\tilde{\Gamma}$ for fixed $\tilde{\sigma}_{\text{s}}^2=\SI{0}{dB}$, while the right panel shows its dependence on the normalized sensing noise variance~$\tilde{\sigma}_{\mathrm{s}}^2$ for fixed $\tilde{\Gamma}=1.2$. The size of each constellation point indicates its probability.}
            \label{fig:results:adaptive_constellations}
    \vspace{-0.3cm}
\end{figure}

\setulcolor{blue}\RevC{\mbox{Fig.~\ref{fig:results:adaptive_constellations}} shows the geometrically and probabilistically shaped constellations for the \mbox{\ac{WF}} generated by the optimized adaptive constellation family for various \mbox{$\tilde{\Gamma}$} and \mbox{$\tilde{\sigma}^2_{\text{s}}$}. We observe that the constellation smoothly transitions from a unit-modulus constellation to a Gaussian-like distribution as \mbox{$\tilde{\Gamma}$} increases and avoids small amplitudes for decreasing \mbox{$\tilde{\sigma}^2_{\text{s}}$.}}

\begin{figure}[!t]
    \centering
    \input{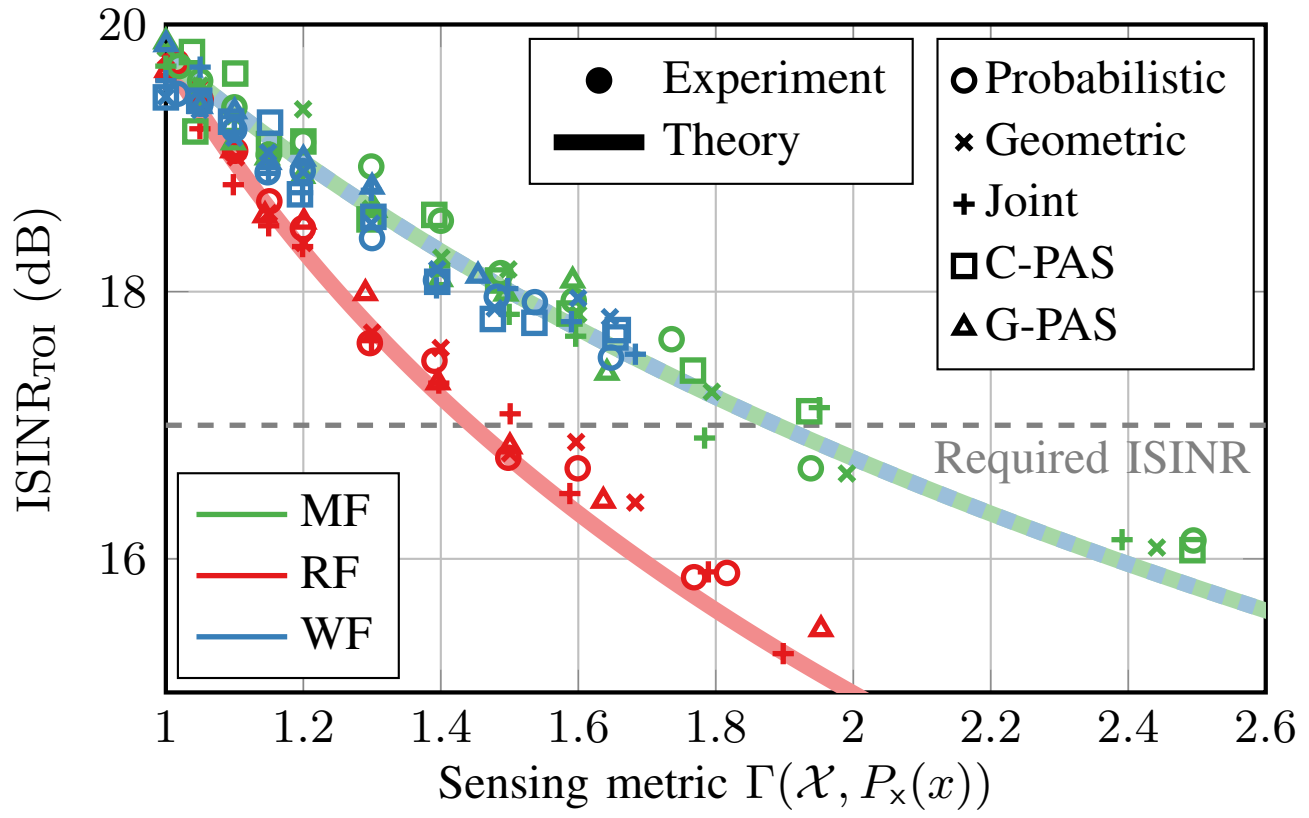}
    \vspace{-0.85cm}
    \caption{Comparison of experimental (markers) and theoretical (curves) \ac{ISINR} for all \acp{SCE} (color) and constellation shaping methods (marker style). The sensing metric~$\Gamma(\mathcal{X}, P_{\RV{x}}(x))$ (x-axis) depends on the \ac{SCE} and is given in Tab.~\ref{tab:theory:Sensing_constraint}.}
    \label{fig:results:sensing_performance_verification}
\end{figure}

\subsubsection{Sensing Performance}
To further validate the derived \ac{ISINR} expressions in Tab.~\ref{tab:ch3:Image_SINR} for a continuous sensing metric~$\Gamma(\mathcal{X}, P_{\RV{x}}(x))$, we transmit optimized constellations with varying sensing metric~$\Gamma(\mathcal{X}, P_{\RV{x}}(x))$ in the previously described two target scenario and measure the \ac{ISINR} of the \ac{TOI}. Fig.~\ref{fig:results:sensing_performance_verification} shows the \ac{ISINR} as a function of the sensing metric~$\Gamma(\mathcal{X}, P_{\RV{x}}(x))$ for all three \acp{SCE} and shaping methods.
Note that the shaping metric $\Gamma(\mathcal{X}, P_{\RV{x}}(x))$ on the x-axis differs across \acp{SCE} according to Tab.~\ref{tab:theory:Sensing_constraint}. The experimental results (markers) closely follow the theoretical predictions, confirming the derivations in Sec.~\ref{sec:theory:SCE}. As exploited during the optimization Sec.~\ref{sec:optimization}, the \ac{ISINR} decreases monotonically with increasing sensing metric~$\Gamma(\mathcal{X}, P_{\RV{x}}(x))$. Moreover, for each \ac{SCE}, all shaping methods follow the same curve, demonstrating that the sensing performance is governed solely by the sensing metric~$\Gamma(\mathcal{X}, P_{\RV{x}}(x))$ (given in Tab.~\ref{tab:theory:Sensing_constraint}) and is independent of the specific shaping method. Assuming a required \ac{ISINR} of \SI{17}{dB} for reliable detection, Fig.~\ref{fig:results:sensing_performance_verification} shows that the detectability of the \ac{TOI} depends on the sensing metric~$\Gamma(\mathcal{X}, P_{\RV{x}}(x))$ and, consequently, on the constellation.

\begin{figure*}[!t]
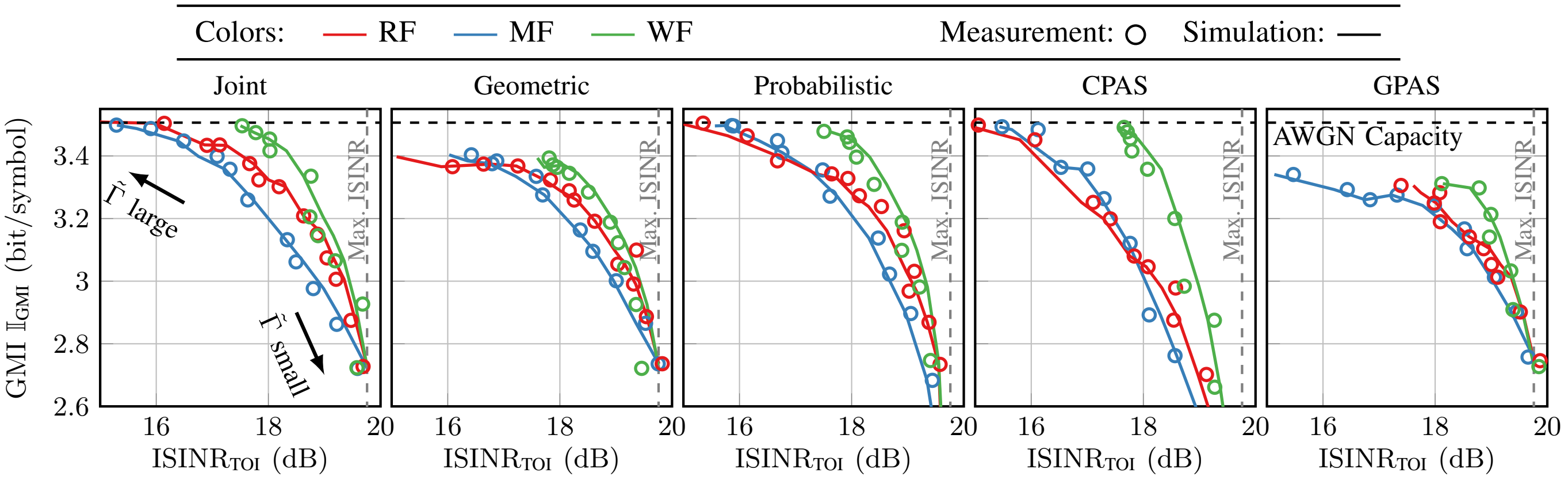

    \vspace{-0.6cm}
    \vspace{-0.4cm}
    \centering
    \begin{tabular}{c c c c p{2cm} c c }
        \toprule
         Colors: &  \begin{tikzpicture}[baseline=-0.5ex] \draw[thick, Set1-A] (0,0) -- (0.5,0);\end{tikzpicture} RF &  \begin{tikzpicture}[baseline=-0.5ex] \draw[thick, Set1-B] (0,0) -- (0.5,0);\end{tikzpicture} MF & \begin{tikzpicture}[baseline=-0.5ex] \draw[thick, Set1-C] (0,0) -- (0.5,0);\end{tikzpicture} WF & \hspace{2cm}&  Measurement: \begin{tikzpicture}[baseline=-0.5ex] \draw [black, thick] (0,0) circle (3pt);\end{tikzpicture} & Simulation: \begin{tikzpicture}[baseline=-0.5ex] \draw[thick, black] (0,0) -- (0.5,0);\end{tikzpicture} \\
         \bottomrule
         \vspace{-2.3cm}
    \end{tabular}
    \begin{picture}(\textwidth,\textwidth/3)
        \put(-0.0035\textwidth,0){\input{figures/ISAC_trade_off/Joint}}
        \put(0.235\textwidth,0){\input{figures/ISAC_trade_off/Geo}}
        \put(0.42\textwidth,0){\input{figures/ISAC_trade_off/Prob}}
        \put(0.605\textwidth,0){\input{figures/ISAC_trade_off/CPAS}}
        \put(0.790\textwidth,0){\input{figures/ISAC_trade_off/GPAS}}
    \end{picture}
    \vspace{-0.8cm}
    \caption{Experimental and simulated \ac{SC} trade-off across all shaping methods and \acp{SCE}. To reduce fluctuations, the theoretical \ac{ISINR} is used on the x-axis.}
    \label{fig:results:ISAC_trade_off_all}
\end{figure*}

\begin{figure}[!b]
    \vspace{-0.6cm}
    \centering
    \input{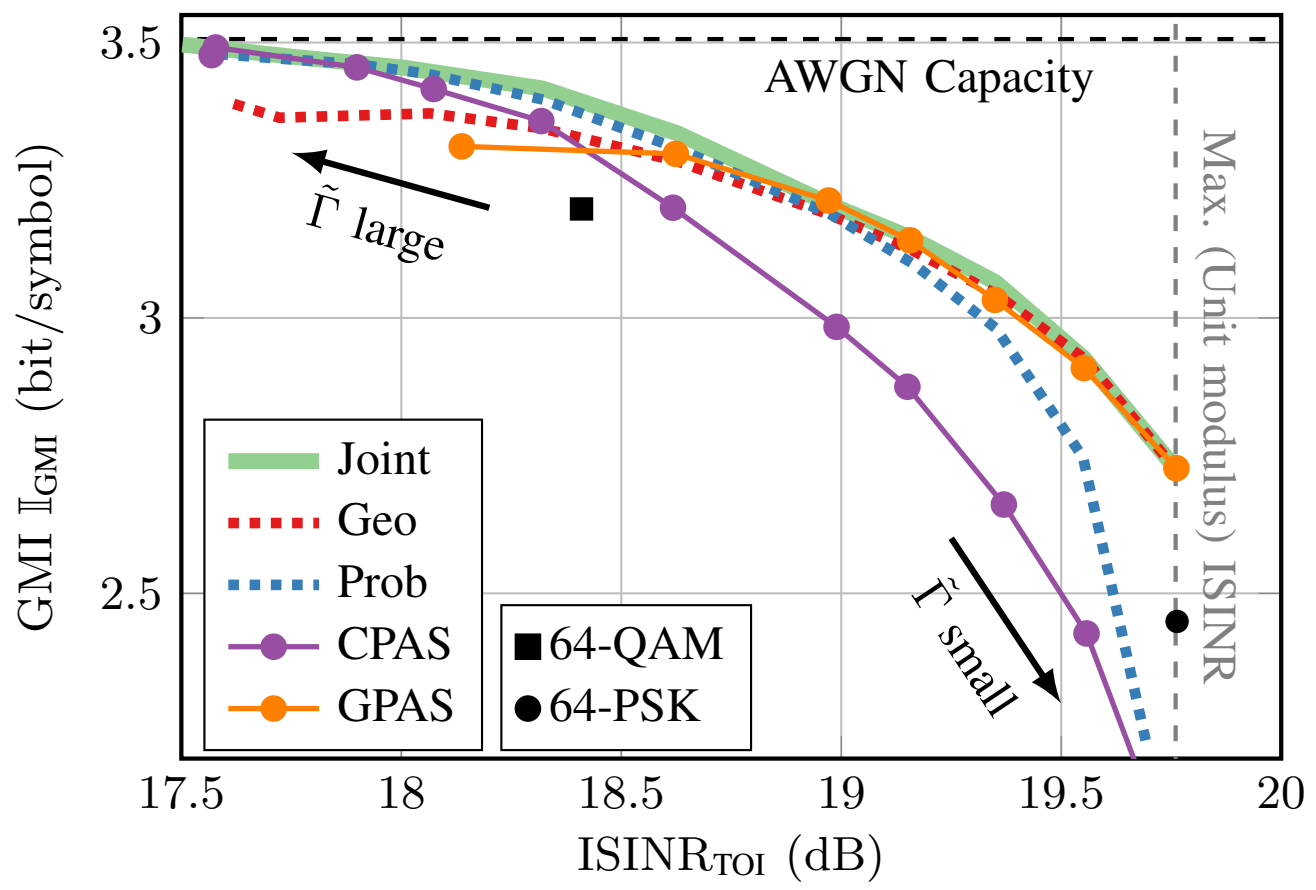}
    \vspace{-0.9cm}
    \caption{\ac{SC} trade-off across shaping methods for the \ac{WF} with the experimentally obtained \ac{GMI}. To reduce fluctuations, the theoretical \ac{ISINR} is used on the x-axis.}
    \label{fig:results:ISAC_trade_off_WF}
\end{figure}

\subsubsection{Communications Performance and \ac{ISAC} Trade-off}
Next, we analyze the communications performance in terms of \ac{GMI} as a function of the sensing performance quantified by the \ac{ISINR} of the \ac{TOI}. To reduce the experimental fluctuations of the \ac{ISINR}, we use the theoretical \ac{ISINR} on the x-axis and the \ac{GMI} obtained from experimental measurements and simulations on the y-axis, which are in good agreement. Fig.~\ref{fig:results:ISAC_trade_off_all} compares the resulting \ac{SC} trade-off for different \acp{SCE} across all shaping methods. As expected, decreasing the sensing threshold~$\tilde{\Gamma}$ improves the sensing performance at the expense of communications performance, thereby verifying the fundamental \ac{ISAC} trade-off.

As discussed in Sec.~\ref{sec:results:SCE}, the achievable \ac{ISINR} depends strongly on the sensing scenario, e.g., low vs. high sensing \ac{SNR} regime. Consequently, either the \ac{MF} or the \ac{RF} may be preferable depending on the scenario, and no general preference should be inferred from Fig.~\ref{fig:results:ISAC_trade_off_all}. In contrast, for a fixed communications performance, the \ac{WF} achieves the highest sensing performance across all shaping methods.

We next consider the extreme points of the trade-off. At the sensing-optimal point, corresponding to the smallest sensing threshold~$\tilde{\Gamma}$, all shaping methods converge to the same \ac{ISINR}, since the optimized constellations approach a unit-modulus constellation. However, the associated communications performance differs across the shaping methods: the \ac{GMI} falls below $\SI{2.6}{bit/symbol}$ for probabilistic shaping and \ac{CPAS}, whereas geometric shaping, \ac{GPAS}, and joint shaping achieve a higher \ac{GMI} of approximately $\SI{2.7}{bit/symbol}$. This difference arises because probabilistic shaping with grid-based constellations is less effective in approximating unit-modulus constellations~\cite{du_probabilistic_2025}.

At the communications-optimal point, corresponding to the largest sensing threshold~$\tilde{\Gamma}$, the sensing constraint becomes inactive and no longer limits the optimization. Consequently, the communications performance depends solely on the shaping method. Geometric shaping and \ac{GPAS} saturate at around $\SI{3.4}{bit/symbol}$, whereas probabilistic shaping, joint shaping, and \ac{CPAS} approach the \ac{AWGN} capacity for large sensing constraints effectively closing the shaping gap~\cite{steiner_comparison_2017}. In this regime, the sensing performance differs across the \ac{SCE}, with the \ac{WF} achieving the highest \ac{ISINR} due to its robustness against noise enhancement and sidelobes for non-unit-modulus constellations.

\begin{figure*}[!b]
    \setcounter{figure}{15}
    \vspace{-0.85cm}
    \vspace{-3cm}
    \centering
    \begin{picture}(\textwidth,\textwidth/2)
        \put(0\textwidth,0){\input{figures/3D_plot_Sens_SD}}
        \put(0.525\textwidth,0){\input{figures/3D_plot_Comm_SD}}
    \end{picture}
    \vspace{-1.0cm}
    \caption{\ac{SC} trade-off across sensing and communications scenarios for the \ac{RF} under an inverse second moment threshold: \ac{ISINR} as a function of the normalized target power and inverse second moment (left), and \ac{GMI} as a function of communications \ac{SNR} and inverse second moment (right).}
    \label{fig:results:interplay:RF}
\end{figure*}

Overall, these results demonstrate that the sensing performance is primarily determined by the choice of the \ac{SCE}, whereas communications performance is governed mainly by the shaping method. Accordingly, the combination of joint shaping and the \ac{WF} achieves the best \ac{ISAC} performance, as the \ac{WF} provides the highest \ac{ISINR}, while joint shaping offers the greatest flexibility for maximizing the \ac{GMI}.

To facilitate a direct comparison across the shaping methods, Fig.~\ref{fig:results:ISAC_trade_off_WF} compares the obtained \ac{SC} trade-off for the \ac{WF}, using the experimentally obtained \ac{GMI}. Our optimized constellations outperform legacy \num{64}-\ac{QAM} and \ac{PSK} in both \ac{SC} performance. Geometric and probabilistic shaping are favorable in sensing-centric and communications-centric regimes, respectively. Joint constellation shaping achieves the best overall performance but increases implementation complexity. \RevB{In particular, joint shaping provides a clear gain over probabilistic shaping
under sensing-constrained operation.}

\begin{figure}[!t]
    \setcounter{figure}{14}
            \centering
            \input{figures/Adaptive_Random_Deterministic}
            \vspace{-0.4cm}
            \caption{\ac{SC} trade-off of the adaptive constellation family and constellations individually optimized for fixed sensing thresholds, considering probabilistic shaping for the \ac{WF}. The adaptive constellation family is optimized over $\tilde{\sigma}_{\text{s}}^2\in[\SI{-5}{dB},\SI{5}{dB}]$, and all constellations are evaluated at $\widetilde{\sigma}_{\mathrm{s}}^2=\SI{-3}{dB}$.}
            \label{fig:results:adaptive:RD}
    \vspace{-0.6cm}
\end{figure}

Importantly, the performance of joint shaping can be closely approached by low-complexity \ac{CPAS} and \ac{GPAS}. In particular, \ac{GPAS} is preferable in the sensing-centric regime (small~$\tilde{\Gamma}$), while \ac{CPAS} excels in the communications-centric regime (large~$\tilde{\Gamma}$). Similar trends across shaping methods are observed for the \ac{RF} and \ac{MF}, and the corresponding results are omitted for brevity. These results demonstrate that the near-joint shaping \ac{ISAC} performance can be achieved with low-complexity shaping methods, making \ac{CPAS} and \ac{GPAS} attractive practical solutions across all considered \acp{SCE}.

\RevC{Next, we compare the \mbox{\ac{SC}} trade-off of the proposed adaptive constellation family with constellations individually optimized for specific sensing metric \mbox{thresholds~$\tilde{\Gamma}$} using our \mbox{\ac{AE}} framework and the \mbox{\acf{MBA}} \mbox{algorithm~\cite{du_probabilistic_2025,du_reshaping_2023}}. To enable a fair comparison \mbox{with~\cite{du_probabilistic_2025}}, we restrict the optimization to probabilistic constellation shaping. \mbox{Fig.~\ref{fig:results:adaptive:RD}} shows that the proposed adaptive constellations achieve \mbox{\ac{MI}} and \mbox{\ac{GMI}} comparable to the individually optimized \mbox{\ac{AE}}-based constellations, while outperforming those optimized using the \mbox{\ac{MBA}} \mbox{algorithm~\cite{du_probabilistic_2025}}. Hence, the proposed adaptive constellation family retains the performance of individually \mbox{\ac{AE}}-optimized constellations while enabling adaptation to varying \mbox{\ac{SC}} requirements. The observed gain over the \mbox{\ac{MBA}}-optimized constellations does not result from a suboptimality of the \mbox{\ac{MBA}} algorithm itself, but from the enlarged design space of our \mbox{\ac{AE}} framework: the constellation amplitudes are rescaled according to the optimized symbol probabilities to satisfy the average power constraint, whereas the \mbox{\ac{MBA}} approach optimizes the probabilities for fixed constellation \mbox{amplitudes~\cite[Alg.~1]{du_reshaping_2023}}.}

\begin{figure*}[!t]
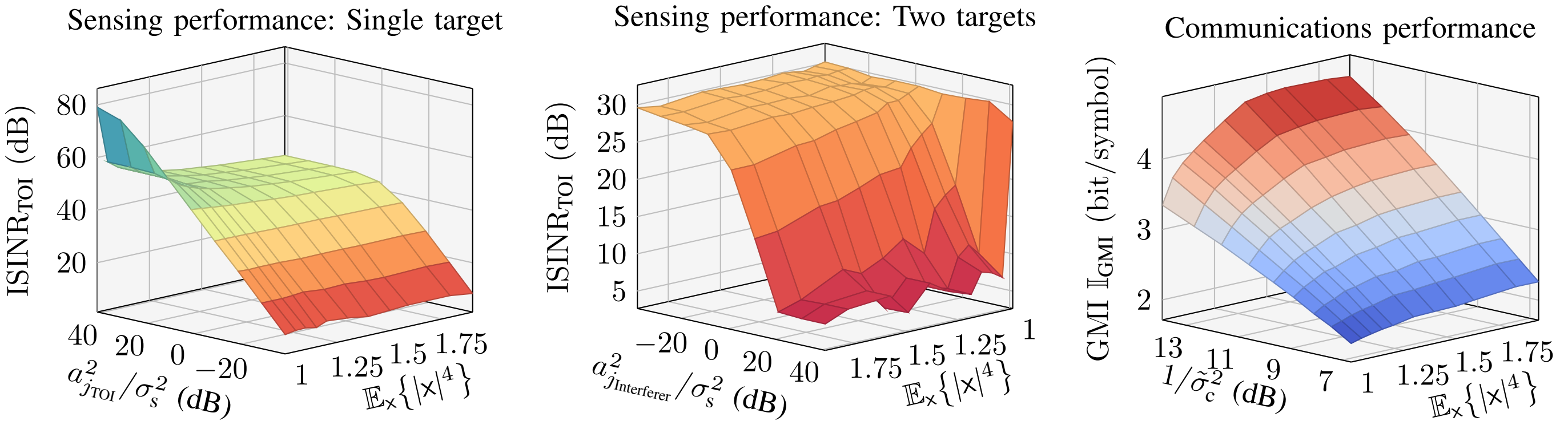

    \setcounter{figure}{16}
    \vspace{-1.0cm}
    \centering
    \begin{picture}(\textwidth,\textwidth/3)
        \put(0\textwidth,0){\input{figures/3D_plot_Sens_MF_ST}}
        \put(0.345\textwidth,0){\input{figures/3D_plot_Sens_MF_MT}}
        \put(0.69\textwidth,0){\input{figures/3D_plot_Comm_MF}}
    \end{picture}
    \vspace{-0.75cm}
    \caption{\ac{SC} trade-off across sensing and communications scenarios for the \ac{MF} under a kurtosis threshold: \ac{ISINR} of the \ac{TOI} as a function of the normalized \ac{TOI}/interfering target power and inverse second moment for single and two target scenario (left and center), and \ac{GMI} as a function of communications \ac{SNR} and kurtosis (right). The same colormaps as in Fig.~\ref{fig:results:interplay:RF} are used.}
    \label{fig:results:interplay:MF}
    \vspace{-0.3cm}
\end{figure*}

\vspace{-0.3cm}
\section{S\&C Trade-Offs Across Scenarios and Constellation Shaping: Experimental Results}
\vspace{-0.2cm}
In this section, we examine the combined effects of the sensing threshold and the communications and sensing channels on the \ac{SC} performance for \acp{SCE}, revealing the resulting \ac{SC} trade-off across a wide range of operating regimes. Fig.~\ref{fig:results:interplay:RF} and~\ref{fig:results:interplay:MF} show the \ac{ISINR} versus the sensing threshold and normalized target power, as well as the \ac{GMI} versus the sensing threshold and communications \ac{SNR}, for the \ac{RF} and \ac{MF}, respectively. For the \ac{RF}, only the single target case is considered, since sensing performance is limited solely by noise enhancement, whereas for the \ac{MF}, both single and two target scenarios are included due to interference depending on the targets. Note that we study only the \ac{RF} and \ac{MF} because these two represent the two fundamental constraint types, and the \ac{WF} achieves similar performance to the superior one in each scenario.

For the \ac{RF}, the \ac{ISINR} increases linearly with the normalized target power. Increasing the inverse second moment $\expecv{\RV{x}}{|\RV{x}|^{-2}}$ introduces a target-independent multiplicative \ac{ISINR} loss, confirming that random signaling affects sensing performance solely through noise enhancement~\eqref{eq:theory:SCE:RF:noise_enhancement}, with low sensing \ac{SNR} operation being particularly critical for detection. The \ac{GMI} increases with both the communications \ac{SNR} and the inverse second moment threshold. For all communications \ac{SNR} values, the \ac{GMI} increases steeply at small thresholds and saturates for larger ones, with the absolute variation in \ac{GMI} increasing with communications \ac{SNR}. 

For the \ac{MF} in the single target scenario, the \ac{ISINR} increases linearly with the normalized target power in the noise-limited regime and is effectively independent of the constellation kurtosis. As the target power increases, interference becomes dominant and the \ac{ISINR} saturates at a kurtosis-dependent level given by~\eqref{eq:theory:SCE:MF:Image_SINR_saturation}, with the strongest sensitivity to kurtosis occurring at low kurtosis values. In the two target case, the interfering target (non-\ac{TOI}) introduces kurtosis-dependent interference, causing the \ac{ISINR} of the \ac{TOI} to become interference-limited at large normalized powers of the interfering target. Consequently, in multi target scenarios, the interfering target can mask the \ac{TOI}, potentially leading to missed detections. The communications performance under the \ac{MF} follows trends similar to the \ac{RF}, with the \ac{GMI} increasing with both communications \ac{SNR} and kurtosis and saturating at large threshold values.

Comparing both estimators reveals distinct \ac{SC} trade-offs. The \ac{MF} exhibits a strongly sensing scenario-dependent \ac{SC} trade-off: random signaling has negligible impact on the sensing performance in the noise-limited regime and thus induces barely any \ac{SC} trade-off, but becomes critical in the interference-limited regime. In contrast, for the \ac{RF}, random signaling introduces sensing scenario-independent noise enhancement, resulting always in a \ac{SC} trade-off. For both estimators, the \ac{SC} trade-off becomes more pronounced with increasing communications \ac{SNR} and is most sensitive to changes for small sensing threshold~$\tilde{\Gamma}$ values.

\section{Conclusion}
\vspace{-0.1cm}
In this work, we optimized and systematically compared geometric, probabilistic (general, \ac{CPAS}, and \ac{GPAS}), and joint constellation shaping for OFDM-\ac{ISAC} systems using a bitwise autoencoder framework, maximizing the \ac{GMI} for the \ac{MF}, \ac{RF}, and \ac{WF} as \acp{SCE}. The resulting \ac{SC} trade-offs were validated through extensive simulations and hardware-in-the-loop experiments. We showed that the \ac{ISINR} depends on second-order statistics of the frequency-domain channel estimate, yielding \ac{SCE}-specific and constellation-dependent \ac{ISINR} expressions.

Our results show that sensing performance is fundamentally governed by how the \ac{SCE} handles transmit signal randomness. For the \ac{RF}, small amplitude symbols cause noise enhancement, leading to \ac{SNR}-independent \ac{ISINR} losses governed by the inverse second-order moment. The \ac{MF} is primarily sensitive to amplitude variations, with sensing degradation scaling with constellation kurtosis and saturating at high sensing \ac{SNR}, whereas the \ac{WF} balances interference and noise enhancement in an \ac{SNR}-dependent manner and achieves the highest \ac{ISINR}. These \ac{SCE}-specific mechanisms directly define the constellation design rules, e.g., reducing amplitude variability for the \ac{MF}, and suppressing small amplitude symbols for the \ac{RF}.

Geometric shaping and \ac{GPAS} achieve a higher \ac{GMI} under strict sensing constraints, while probabilistic shaping and \ac{CPAS} perform better under relaxed sensing constraints. Notably, joint shaping combines the strengths of both geometric and probabilistic constellation shaping, significantly outperforming legacy constellation formats and offering a flexible \ac{SC} trade-off. Consequently, sensing performance is primarily determined by the \ac{SCE}, while communications performance is dominated by the shaping method.

From a system design perspective, joint constellation shaping combined with the \ac{WF} achieves the best overall \ac{ISAC} trade-off by exploiting both geometric and probabilistic degrees of freedom. At the same time, low-complexity \ac{CPAS} and \ac{GPAS} closely approach joint-shaping performance, offering a practical performance-complexity trade-off for implementation-constrained \ac{ISAC} systems. By systematically characterizing the \ac{SC} trade-off across \acp{SCE}, shaping strategies, and operating regimes, this work bridges theoretical analysis and practical system design in \ac{OFDM}-\ac{ISAC} systems.

\setulcolor{red}\RevC{Complementary over-the-air measurements in uncontrolled propagation environments constitute an important direction for future research, allowing the impact of environmental clutter, multipath, and other propagation effects to be further investigated.} \RevA{Another direction for future work is the investigation of constellation shaping for transceivers operated in their nonlinear regime.} %

\vspace{-0.3cm}
\section*{Acknowledgments}
\vspace{-0.05cm}
The authors would like to thank Rohde \& Schwarz for providing the AREG800A for the measurements.

\appendices
\vspace{-0.2cm}
\section{}
\vspace{-0.1cm}
\label{ch:appendix:target_power}
To derive the deterministic target power in the sensing image~(\ref{eq:theory:target_power}), we successively insert the \ac{RDM} definition~(\ref{eq:system_model:RDM_calculation}), the windowing operation~(\ref{eq:system_model:windowing}), and the decomposition of the sensing channel estimate~(\ref{eq:theory:channel_estiate_seperability}) into~(\ref{eq:theory:target_power}).
\vspace{-0.2cm}
\begin{align}
    & \left| \expecv{\RV{x},\RV{w}^{\text{s}}}{\hat{\RV{h}}^{\text{win}}_{k_{\text{TOI}},p_{\text{TOI}}}} \right|^2 \notag \\
    & = \!  \left| \!\frac{1}{\sqrt{MN}} \! \sum_{m=0}^{M-1} \sum_{n=0}^{N-1} \!  \alpha h_{n,m} u_{n,m} \e^{\j 2 \pi \frac{n}{\eta_\tau N}k_{\text{TOI}}} \e^{-\j 2 \pi \frac{m}{\eta_\text{D} M} p_{\text{TOI}}} \!\right|^{2} \label{eq:apdx:target:a} \\[-12pt]
    & = \frac{\alpha^2}{MN} \Bigg| \sum_{m=0}^{M-1} \sum_{n=0}^{N-1} \biggl( a_{j_\text{TOI}} \e^{\j \varphi_{j_\text{TOI}}}u_{n,m} \nonumber \\[-9pt]
    & + \sum_{j \neq j_\text{TOI}}^{J} a_j \e^{\j \varphi_j} u_{n,m} \e^{\j 2 \pi \frac{n}{N} (\frac{k_{\text{TOI}}}{\eta_\tau} -\tilde{\tau}_j)} \e^{-\j 2 \pi \frac{m}{M} (\frac{p_{\text{TOI}}}{\eta_\text{D}} - \tilde{f}_{\text{D,}j})} \biggr) \Bigg|^2 \label{eq:appendix:target_power_derivation_b} \\[-15pt]
    & \approx \frac{\alpha^2 a^2_{j_\text{TOI}}}{MN} \left( \sum_{m=0}^{M-1} \sum_{n=0}^{N-1} u_{n,m} \right)^2
\end{align} %
In~(\ref{eq:apdx:target:a}), we use that taking the expectation removes the zero mean noise-and-interference component $\tilde{\RV{w}}^{\text{s}}_{n,m}$. Next, we insert the sensing channel model~(\ref{eq:system_model:sensing_channel_frequency_domain}). For sufficiently large oversampling factors $\eta_\tau$ and $\eta_\text{D}$, the delay and Doppler of the \ac{TOI} align with the corresponding delay-Doppler bin, i.e., \mbox{$k_{\text{TOI}}/\eta_\tau = \tilde{\tau}_{j_\text{TOI}}$} and \mbox{$p_{\text{TOI}}/\eta_\text{D} = \tilde{f}_{\text{D},j_\text{TOI}}$}, which allows us to separate the \ac{TOI} contribution from the remaining targets in~\eqref{eq:appendix:target_power_derivation_b}. The second term in~\eqref{eq:appendix:target_power_derivation_b} corresponds to the two-dimensional (I)\ac{DFT} of the window $u_{n,m}$ evaluated at the delay and Doppler offsets
\mbox{$k_{\text{TOI}}/\eta_\tau - \tilde{\tau}_{j}$} and Doppler offset \mbox{$p_{\text{TOI}}/\eta_\text{D} - \tilde{f}_{\text{D,}j}$}. Since the window $u_{n,m}$~(\ref{eq:system_model:windowing}) is designed to suppress sidelobes and the \ac{TOI} is well separated from all other targets~(\ref{eq:theory:target_separation_condition}), the contribution from the other targets are negligible at the \ac{TOI} bin. Hence, the second term can be neglected, and~\eqref{eq:appendix:target_power_derivation_b} reduces to~\eqref{eq:theory:target_power}. \RevA{If the target separation assumption~(\ref{eq:theory:target_separation_condition}) is violated, adjacent targets leak coherently into the \ac{TOI} bin, biasing the power in the \ac{TOI} bin depending on the relative phase and rendering the detection probability~\eqref{eq:pd_isinr} inapplicable.}

\section{}
\label{ch:appendix:noise-and-interference-power}
\vspace{-0.15cm}
To derive the variance in the sensing image~(\ref{eq:theory:noise-and-interference-power}), we successively insert the \ac{RDM} definition~(\ref{eq:system_model:RDM_calculation}), the windowing operation~(\ref{eq:system_model:windowing}), and the decomposition of the sensing channel estimate~(\ref{eq:theory:channel_estiate_seperability}) into~(\ref{eq:theory:noise-and-interference-power}).
\vspace{-0.2cm}
\begin{align}
    &\mathrm{Var}_{\RV{x}, \RV{w}^{\text{s}}} \!\left\{ \hat{\RV{h}}^{\text{win}}_{k,p} \right\} \notag \\[-4pt]
    &= \!\mathrm{Var}_{\RV{x}, \RV{w}^{\text{s}}}\! \left\{ \!
        \frac{1}{\sqrt{MN}} \!\!
        \sum_{m=0}^{M-1} \sum_{n=0}^{N-1} \!
        u_{n,m} \tilde{\RV{w}}^{\text{s}}_{n,m}
        \e^{\j 2 \pi \frac{n}{\eta_\tau N}k}
        \e^{-\j 2 \pi \frac{m}{\eta_\text{D} M} p}
    \right\} \label{eq:apdx:noise_power:a} \\[-15pt]
    &=  \frac{1}{MN}
        \sum_{m=0}^{M-1} \sum_{n=0}^{N-1}
        u^2_{n,m} \,
        \mathrm{Var}_{\RV{x}, \RV{w}^{\text{s}}} \! \left\{ \tilde{\RV{w}}^{\text{s}}_{n,m} \right\} \label{eq:apdx:noise_power:b} \\
    &\approx  \frac{1}{MN}
        \sum_{m=0}^{M-1} \sum_{n=0}^{N-1}
        \mathrm{Var}_{\RV{x}, \RV{w}^{\text{s}}} \! \left\{ \hat{\RV{h}}_{n,m} \right\}.
        \label{eq:appendix:noise_power_derivation_b}
\end{align}
We exploit in~\eqref{eq:apdx:noise_power:a} that the deterministic part does not affect the variance, and in~\eqref{eq:apdx:noise_power:b} that the noise-and-interference samples $\tilde{\RV{w}}^{\text{s}}_{n,m}$ are uncorrelated. Finally, we exploit in~\eqref{eq:appendix:noise_power_derivation_b} that the window is normalized to unit average power, has a negligible correlation with the variance $\mathrm{Var}_{\RV{x}, \RV{w}^{\text{s}}}\! \left\{ \tilde{\RV{w}}^{\text{s}}_{n,m} \right\}$ and that the variance of the noise-and-interference equals the variance in the frequency-domain channel estimate~(\ref{eq:system_model:frequency_domain_channel_estimate}), yielding~\eqref{eq:theory:noise-and-interference-power}.

\vspace{-0.4cm}
\section{}
\vspace{-0.15cm}
\label{sec:appendix:entropy_maximization}
In this appendix, we show that solving the nonlinear system~(\ref{eq:bound:non_linear_eq_system}) is equivalent to solving~(\ref{eq:theory:Cm2}), (\ref{eq:theory:C0}), and (\ref{eq:theory:C2}). We begin by rewriting~(\ref{eq:bound:non_linear_eq_system}) for \mbox{$c \in \{ -1, 0, 1\}$} as
\vspace{-0.2cm}
\begin{align}
    C_{2c} & = \int_0^{2\pi} \int_0^\infty r^{2c+1} \mathrm{e}^{\gamma^{\text{RF}}_{0} + \gamma^{\text{RF}}_{2} r^2 + \gamma^{\text{RF}}_{-2}/r^2} \,\mathrm{d}r\,\mathrm{d}\theta \label{eq:appdx:a} \\[-3pt]
     & = 2\pi \e^{\gamma^{\text{RF}}_{0}} \int_0^\infty r^{2c+1} \e^{\gamma^{\text{RF}}_{2} r^2 + \gamma^{\text{RF}}_{-2}/r^2} \,\mathrm{d}r \label{eq:appdx:b} \\
     & = \pi \e^{\gamma^{\text{RF}}_{0}} \int_0^\infty u^{c} \e^{\gamma^{\text{RF}}_{2} u + \gamma^{\text{RF}}_{-2} /u} \,\mathrm{d}u, \label{eq:appdx:c}
\end{align}
where we transform a 2D Cartesian complex-plane integral to polar coordinates with \mbox{$x = r \e^{\j \theta}$} in~(\ref{eq:appdx:a}), and substitute \mbox{$u = r^2$} in~(\ref{eq:appdx:c}). Using the integral identity~\cite[entry:~3.471.9]{gradshteyn2007}, we obtain~(\ref{eq:theory:Cm2}), (\ref{eq:theory:C0}), and (\ref{eq:theory:C2}).


\vspace{-0.4cm}
\begin{thebibliography}{10}
\vspace{-0.15cm}
\providecommand{\url}[1]{#1}
\csname url@samestyle\endcsname
\providecommand{\newblock}{\relax}
\providecommand{\bibinfo}[2]{#2}
\providecommand{\BIBentrySTDinterwordspacing}{\spaceskip=0pt\relax}
\providecommand{\BIBentryALTinterwordstretchfactor}{4}
\providecommand{\BIBentryALTinterwordspacing}{\spaceskip=\fontdimen2\font plus
\BIBentryALTinterwordstretchfactor\fontdimen3\font minus \fontdimen4\font\relax}
\providecommand{\BIBforeignlanguage}[2]{{%
\expandafter\ifx\csname l@#1\endcsname\relax
\typeout{** WARNING: IEEEtran.bst: No hyphenation pattern has been}%
\typeout{** loaded for the language `#1'. Using the pattern for}%
\typeout{** the default language instead.}%
\else
\language=\csname l@#1\endcsname
\fi
#2}}
\providecommand{\BIBdecl}{\relax}
\BIBdecl

\bibitem{liu_integrated_2022}
F.~Liu \emph{et~al.}, ``Integrated sensing and communications: Toward dual-functional wireless networks for {6G} and beyond,'' \emph{{IEEE} J. Sel. Areas Commun.}, vol.~40, no.~6, pp. 1728--1767, Jun. 2022.

\bibitem{wild_6g_2023}
T.~Wild, A.~Grudnitsky, S.~Mandelli, M.~Henninger, J.~Guan, and F.~Schaich, ``{6G} integrated sensing and communication: From vision to realization,'' in \emph{Proc. European Radar Conference {(EuRAD)}}, Berlin, Germany, Sep. 2023, pp. 355--358.

\bibitem{keskin_fundamental_2025}
M.~F. Keskin \emph{et~al.}, ``Fundamental trade-offs in monostatic {ISAC}: A holistic investigation towards {6G},'' \emph{{IEEE} Trans. Wireless Commun.}, vol.~24, no.~9, pp. 7856--7873, Apr. 2025.

\bibitem{du_probabilistic_2025}
Z.~Du \emph{et~al.}, ``Probabilistic constellation shaping for {OFDM} {ISAC} signals under temporal-frequency filtering,'' \emph{{IEEE} Trans. Wireless Commun.}, vol.~25, pp. 12\,914--12\,929, 2026.

\bibitem{han_constellation_2025}
K.~Han, K.~Meng, A.~Chatzicharistou, and C.~Masouros, ``Constellation design in {OFDM}-{ISAC} over data payloads: From {MSE} analysis to experimentation,'' Oct. 2025, preprint, available at \url{https://arxiv.org/abs/2510.13101}.

\bibitem{xu_experimental_2024}
J.~Xu, Z.~Du, J.~Wang, and Y.~Xu, ``An experimental validation of {ISAC} with probabilistic constellation shaping under {OFDM} signaling,'' in \emph{Proc. {IEEE} Int. Conf. on Unmanned Systems ({ICUS})}, Nanjing, China, Oct. 2024, pp. 1579--1584.

\bibitem{xiong_torch_2024}
Y.~Xiong, F.~Liu, K.~Wan, W.~Yuan, Y.~Cui, and G.~Caire, ``From torch to projector: Fundamental tradeoff of integrated sensing and communications,'' \emph{IEEE BITS the Inf. Theory Mag.}, vol.~4, no.~1, pp. 73--90, Mar. 2024.

\bibitem{stark_joint_2019}
M.~Stark, F.~Ait~Aoudia, and J.~Hoydis, ``Joint learning of geometric and probabilistic constellation shaping,'' in \emph{Proc. {IEEE} {GLOBECOM} {Workshops}}, Waikoloa, HI, USA, Dec. 2019.

\bibitem{rode_end--end_2023}
A.~Rode, B.~Geiger, S.~Chimmalgi, and L.~Schmalen, ``End-to-end optimization of constellation shaping for {Wiener} phase noise channels with a differentiable blind phase search,'' \emph{J. Lightw. Technol.}, vol.~41, no.~12, pp. 3849--3859, Jun. 2023.

\bibitem{du_probabilistic_2023}
Z.~Du \emph{et~al.}, ``Probabilistic constellation shaping for {OFDM}-based {ISAC} signaling,'' in \emph{Proc. Global Commun. Conf. (GLOBECOM)}, Kuala Lumpur, Malaysia, Dec. 2023, pp. 509--514.

\bibitem{du_reshaping_2023}
Z.~Du, F.~Liu, Y.~Xiong, T.~X. Han, Y.~C. Eldar, and S.~Jin, ``Reshaping the {ISAC} tradeoff under {OFDM} signaling: A probabilistic constellation shaping approach,'' \emph{{IEEE} Trans. Signal Process.}, vol.~72, pp. 4782--4797, Sep. 2024.

\bibitem{Geiger25JCS}
B.~Geiger, F.~Liu, S.~Lu, A.~Rode, and L.~Schmalen, ``Joint optimization of geometric and probabilistic constellation shaping for {OFDM}-{ISAC} systems,'' in \emph{Proc. IEEE Int. Symp. on Joint Communications and Sensing (JC\&S)}, Oulu, Finland, Jan. 2025.

\bibitem{geiger_constellation_2025}
B.~Geiger \emph{et~al.}, ``Constellation shaping for {OFDM}-{ISAC} systems: {From} theoretical bounds to practical implementation,'' \emph{{IEEE} Trans. Commun.}, {Early Access,} 2026.
\RevC{\bibitem{cammerer_trainable_2020}
S.~Cammerer, F.~{Ait Aoudia}, S.~Dörner, M.~Stark, J.~Hoydis, and S.~ten Brink, ``Trainable communication systems: Concepts and prototype,'' \emph{{IEEE} Trans. Commun.}, vol.~68, no.~9, pp. 5489--5503, Sep. 2020.}

\bibitem{yang_constellation_2024}
X.~Yang, R.~Zhang, D.~Zhai, F.~Liu, R.~Du, and T.~X. Han, ``Constellation design for integrated sensing and communication with random waveforms,'' \emph{{IEEE} Trans. Wireless Commun.}, vol.~23, no.~11, pp. 17\,415--17\,428, Nov. 2024.

\bibitem{Ivanov_BICM}
M.~Ivanov, C.~Häger, F.~Brännström, A.~Graell~i Amat, A.~Alvarado, and E.~Agrell, ``On the information loss of the max-log approximation in {BICM} systems,'' \emph{{IEEE} Trans. Inf. Theory}, vol.~62, no.~6, pp. 3011--3025, Jun. 2016.

\bibitem{bocherer_bandwidth_2015}
G.~Böcherer, F.~Steiner, and P.~Schulte, ``Bandwidth efficient and rate-matched low-density parity-check coded modulation,'' \emph{{IEEE} Trans. Commun.}, vol.~63, no.~12, pp. 4651--4665, Dec. 2015.

\bibitem{fehenberger_probabilistic_2016}
T.~Fehenberger, A.~Alvarado, G.~Böcherer, and N.~Hanik, ``On probabilistic shaping of quadrature amplitude modulation for the nonlinear fiber channel,'' \emph{J. Lightw. Technol.}, vol.~34, no.~21, pp. 5063--5073, Nov. 2016.

\bibitem{braun_ofdm_2014}
M.~Braun, ``{OFDM} radar algorithms in mobile communication networks,'' Ph.D. dissertation, Karlsruhe Institute of Technology, 2014.
\RevA{\bibitem{kumari_ieee_2018}
P.~Kumari, J.~Choi, N.~González-Prelcic, and R.~W. Heath, ``{IEEE} 802.11ad-based radar: An approach to joint vehicular communication-radar system,'' \emph{{IEEE} Trans. Veh. Technol.}, vol.~67, no.~4, pp. 3012--3027, Apr. 2018.}
\RevA{\bibitem{liu_uncovering_2025}
F.~Liu \emph{et~al.}, ``Uncovering the iceberg in the sea: {Fundamentals} of pulse shaping and modulation design for random {ISAC} signals,'' \emph{{IEEE} Trans. Signal Process.}, vol.~73, pp. 2511--2526, 2025.}

\bibitem{nuss_limitations_2018}
B.~Nuss, J.~Mayer, and T.~Zwick, ``Limitations of {MIMO} and multi-user access for {OFDM} radar in automotive applications,'' in \emph{Proc. {IEEE} Int. Conf. on Microwaves for Intelligent Mobility ({ICMIM})}, 2018.
\RevA{\bibitem{richards_fundamentals_2014}
M.~A. Richards, \emph{\BIBforeignlanguage{en}{Fundamentals of {Radar} {Signal} {Processing}}}, 2nd~ed.\hskip 1em plus 0.5em minus 0.4em\relax New York: McGraw-Hill Education, 2014.}

\bibitem{Cover2006}
T.~Cover and J.~Thomas, \emph{{Elements of Information Theory}}, 2nd~ed.\hskip 1em plus 0.5em minus 0.4em\relax Hoboken, NJ, USA: Wiley, 2006.

\bibitem{Nuss_Testbed}
B.~Nuss \emph{et~al.}, ``Flexible and scalable broadband massive {MIMO} testbed for joint communication and sensing applications,'' in \emph{Proc. European Wireless}, Brno, Czech Republic, Sep. 2024, pp. 187--192.
\RevB{\bibitem{ochiai_distribution_2001}
H.~Ochiai and H.~Imai, ``On the distribution of the peak-to-average power ratio in {OFDM} signals,'' \emph{{IEEE} Trans. Commun.}, vol.~49, no.~2, pp. 282--289, Feb. 2001.}

\bibitem{areg_datasheet}
{Rohde \& Schwarz}, ``{AREG800A Automotive Radar Echo Generator},'' \url{https://scdn.rohde-schwarz.com/ur/pws/dl\_downloads/ pdm/cl\_brochures\_and\_datasheets/product\_brochure/3609\_8015\_12/ AREG800A\_bro\_en\_3609-8015-12\_v0500.pdf}, Sep. 2025, version 09.00, accessed Feb.~2026.
\RevC{\bibitem{diewald_radar_2021}
A.~Diewald \emph{et~al.}, ``Radar target simulation for vehicle-in-the-loop testing,'' \emph{Vehicles}, vol.~3, no.~2, pp. 257--271, Jun. 2021.}

\bibitem{mandelli_survey_2023}
S.~Mandelli, M.~Henninger, M.~Bauhofer, and T.~Wild, ``Survey on integrated sensing and communication performance modeling and use cases feasibility,'' in \emph{Proc. {International} {Conference} on {6G} {Networking} ({6GNet})}, Paris, France, Oct. 2023.

\bibitem{steiner_comparison_2017}
F.~Steiner and G.~Böcherer, ``Comparison of geometric and probabilistic shaping with application to {ATSC} 3.0,'' in \emph{Proc. Int. ITG Conf. on Systems, Communications and Coding (SCC)}, Hamburg, Germany, Feb. 2017.

\bibitem{gradshteyn2007}
I.~S. Gradshteyn and I.~M. Ryzhik, \emph{Table of Integrals, Series, and Products}, 7th~ed.\hskip 1em plus 0.5em minus 0.4em\relax London, UK: Academic Press, 2007.

\end{thebibliography}
\end{document}